%% file: NGC5460_v3.0.tex
\documentclass[useAMS,usenatbib]{mn2e}
\usepackage{txfonts}
\usepackage{longtable}  
\usepackage{natbib}
\usepackage{graphicx}
\usepackage{graphics}
\usepackage{amssymb}
\bibpunct{(}{)}{;}{a}{}{,}
\def\Teff{\ensuremath{T_{\mathrm{eff}}}}
\def\logg{\ensuremath{\log g}}
\def\vmic{$\upsilon_{\mathrm{mic}}$}

\def\vsini{\ensuremath{{\upsilon}\sin i}}
\def\kms{$\mathrm{km\,s}^{-1}$}

\def\exc{$\chi_{\mathrm{excit}}$}

\def\vr{${\upsilon}_{\mathrm{r}}$}
\def\logt{\ensuremath{\log t}}
\def\espa{ESPaDOnS}

\def\nlte{non-LTE}
\def\llm{{\sc LLmodels}}

\def\logl{\ensuremath{\log L/L_{\odot}}}
\def\mbol{$M_{\mathrm{bol}}$}

\def\M{\ensuremath{M/M_{\odot}}}
\def\vald{{\sc VALD}}
\def\synth{{\sc SYNTH3}}
\title[A detailed spectroscopic analysis of the open cluster NGC\,5460]
{A detailed spectroscopic analysis of the open cluster NGC\,5460\thanks{Based 
on observations made with ESO Telescopes at the Paranal Observatory under 
programme ID~079.D-0178A}} 

\author[L. Fossati et al.]{L. Fossati,$^{1,2}$\thanks{l.fossati@open.ac.uk}
			   C.P. Folsom,$^{3}$
			   S. Bagnulo,$^{3}$
			   J.H. Grunhut,$^{4}$
			   O. Kochukhov,$^{5}$
		\newauthor J.D. Landstreet,$^{3,6}$
			   C. Paladini,$^{1}$
			   G.A. Wade,$^{4}$\\
$^{1}$Institut f\"ur Astronomie, Universit\"{a}t Wien, 
T\"{u}rkenschanzstrasse 17, 1180 Wien, Austria\\
$^{2}$Department of Physics and Astronomy, Open University, 
Walton Hall, Milton Keynes MK7 6AA, UK\\
$^{3}$Armagh Observatory, College Hill, Armagh BT61 9DG, Northern Ireland, UK\\
$^{4}$Physics Dept., Royal Military College of Canada, 
PO Box 17000, Station Forces, K7K 4B4, Kingston, Canada\\
$^{5}$Department of Physics and Astronomy, 
Uppsala University, 751 20, Uppsala, Sweden\\
$^{6}$Department of Physics \& Astronomy, 
University of Western Ontario, London, N6A 3K7, Ontario, Canada}
\begin{document}

\date{}

\pagerange{\pageref{firstpage}--\pageref{lastpage}} \pubyear{2010}

\maketitle

\label{firstpage}

\begin{abstract}
Within the context of a large project aimed at studying early F-, A- and 
late B-type stars we present the abundance analysis of the photospheres of 
21 members of the open cluster NGC\,5460, an intermediate age cluster 
(\logt $\sim 8.2$) previously unstudied with spectroscopy. Our study is based 
on medium and high resolution spectra obtained with the FLAMES instrument of 
the ESO/VLT. We show that cluster members have a nearly solar metallicity, 
and that there is evidence that the abundances of magnesium and iron are 
correlated with the effective temperature, exhibiting a maximum around
\Teff=10500\,K. No correlations are found between abundances and projected 
equatorial velocity, except for marginal evidence of barium being more 
abundant in slower than in faster rotating stars. We discovered two He-weak 
stars, and a binary system where the hotter component is a HgMn star. 
We provide new estimates for the cluster distance ($720\pm50$\,pc), age 
(\logt=$8.2\pm0.1$), and mean radial velocity ($-17.9\pm5.2$\,\kms).
\end{abstract}

\section{Introduction}\label{introduction}
The photospheres of main sequence early F-, A- and late B-type stars display 
the signatures of different physical effects of comparable magnitudes,
including large and relatively simple magnetic fields, strong surface 
convection, the presence of emission lines, pulsation, diffusion of chemical 
elements under the competing influences of gravity and radiative acceleration, 
and various kinds of mixing processes, from small-scale turbulence to global 
circulation currents. Detailed studies of these stars are useful to explore 
and understand the physics of these various phenomena, most of which certainly 
play a role in the atmosphere of all the other kinds of stars across the 
Hertzsprung-Russell diagram, but are often more difficult to observe. 

This work is a part of a large project aimed at setting constraints to the 
diffusion theory, by means of a systematic study of the abundances of the 
chemical elements of early F-, A- and late B-type star that are cluster 
members. Cluster members are specially interesting mainly for two reasons: 
1) stars belonging to the same cluster were born from the same initial 
composition, and the comparison of the abundances of the chemical elements 
that compose their atmosphere may help to characterise how diffusion works 
for different stellar masses and effective temperatures; 2) the age of 
cluster members can be determined with much higher accuracy than for field 
stars, hence a systematic study of the abundances of the chemical elements of
F-, A- and B-type stars belonging to different clusters may help to
identify the characteristic time-scales of diffusion.

Using several instruments (FLAMES at ESO/VLT, FIES at the Nordic
Optical Telescope, ELODIE and SOPHIE at the Observatorie de Haute
Provence, and \espa\ at the CHFT), we have obtained low, medium, and
high resolution spectra for about one thousand stars in the fields of
view of a dozen of open clusters with age ranging from \logt=6.8 to
8.9, and with distance modulus between 6.4 and 14.2 
\citep[see][for the full list of observed open clusters]{fossati2008coska}. 
One of our ongoing projects based on these data is to perform accurate 
abundance analysis of early G, F, A, and late B-type cluster members, 
searching for a correlation between age and chemical composition of the 
stellar photospheres. Analysis of individual clusters will be used to search
for correlations between chemical composition, stellar temperature and
stellar rotation. This project was started with a detailed study of the
Praesepe cluster \citep{fossati2007,fossati2008,fossati10}. Similar works
performed by other groups include the analysis of Coma Berenices
\citep{gebran2008}, the Pleiades \citep{Gebetal08}, the Hyades 
\citep{gebran2010} and NGC\,6475 \citep{Viletal09}. The analysis of the open 
cluster NGC\,5460 is the subject of this paper.
\section{Observations}\label{obs}
NGC\,5460 is an intermediate age open cluster
\citep[\logt=8.20$\pm$0.20,][]{ahumada2007} that includes a large number of 
early-type stars, covering a large range in stellar mass (up to about 
4\,$M_{\odot}$). NGC\,5460 is spectroscopically almost completely 
unstudied, except for a few polarimetric observations \citep{bagnulo2006}. 
NGC\,5460 was observed in service mode with FLAMES, the multi-object 
spectrograph attached to the Unit 2 Kueyen of the ESO/VLT, on May 22 2007. 
The FLAMES instrument \citep{pasetal02} is able to access targets over a 
field of view of 25\,arcmin in diameter. With its MEDUSA fibers linked to 
the GIRAFFE spectrograph, FLAMES can acquire low or medium resolution spectra 
($R=7\,500-30\,000$) of up to 130 stars. GIRAFFE low resolution spectra cover 
wavelength ranges of 500 up to 1200\,\AA, while mid resolution spectra cover 
wavelength intervals of 170 -- 500\,\AA, all within the visible range 
3700--9000\,\AA. With the UVES fibers linked to the red arm of the UVES 
spectrograph, FLAMES can simultaneously obtain high resolution spectra 
($R=47\,000$) for up to 8 stars, covering an interval range of about 
2000\,\AA, centred about 5200, or 5800, or 8600\,\AA.
\subsection{Instrument setups}\label{setup}
Taking into account that for the stellar parameter determination it is
valuable to have observations of the hydrogen lines, and that the
number of spectral lines (and the number of chemical elements that may
be analysed) increases towards shorter wavelengths, for FLAMES/UVES we
adopted the 520\,nm setting, which covers the wavelength region
4140--6210\,\AA. The telescope was guided at the central wavelength of
this setting (5200\,\AA).

To minimise light losses due to the atmosphere differential refraction, 
for GIRAFFE we were forced to choose a setting with central wavelength not 
too far from 5200\,\AA. We selected the HR09 setting, which, with a 
resolving power of 25\,900, covers the wavelength region 5139--5355\,\AA.
Spectra obtained with this setting cover lines of a number of Fe-peak
elements, such as iron, titanium and chromium, in addition to the Mg triplet at
$\lambda\lambda\sim$\,5167, 5172 and 5183. UVES and GIRAFFE spectra were 
being taken simultaneously. However, due to its lower spectral resolution, 
the GIRAFFE setting required a substantially shorter exposure time than the 
UVES setting (for a given signal to noise ratio per spectral bin). While UVES 
shutter was still open, the extra-time that was left available for 
observations with GIRAFFE was used to acquire an additional mid resolution 
spectrum, using setting HR11, and a low resolution spectrum, adopting 
setting LR03. The setting HR11 covers the wavelength range 
5592--5838\,\AA\ with a resolving power of 24\,200. Spectra obtained with 
this setting cover iron, sodium and scandium lines, where the latter are 
important to classify A-type stars as Am chemically peculiar stars. Setting 
LR03 covers the spectral range 4500--5077\,\AA\ with a resolving 
power of 7\,500. This setting was selected mainly to cover the H$\beta$
line, used as primary indicator for the atmospheric parameters. As to make 
sure to avoid saturation of the brightest targets, UVES observations were 
divided into four sub-exposures, while for each of the GIRAFFE settings we 
obtained two sub-exposures.

Our data were obtained in service mode, and the package released to us 
included the products reduced by ESO with the instrument dedicated pipelines.  
In this work we used the low and mid-resolution GIRAFFE data as reduced by 
ESO\footnote{see the FLAMES web site 
({\tt http://www.eso.org/sci/facilities/\\paranal/instruments/flames/}) for 
more details on the reduction of GIRAFFE data.}. Due to the high precision 
required for our investigation, we decided to repeat the data reduction of 
the UVES data. For this purpose we used the UVES standard ESO-pipeline
vers.\,2.9.7 developed on the MIDAS platform vers.\,08SEP patch level pl1.1. 
The pipeline included all the main steps of calibration: bias subtraction, 
flat field correction and wavelength calibration.
\subsection{Target selection}\label{targ.selection}
Our target list is the result of three selection processes. 
%--------------------------------------------------------------------
\input{./Table_Observations.tex}
%--------------------------------------------------------------------
The first selection was performed when the FLAMES observing blocks were 
prepared. The position of the Medusa fibers was determined by the
Fiber Positioning System Software (FPOSS), which was fed with an input
catalogue extracted from the UCAC2 catalogue \citep{UCAC2} centred
about (RA DEC) = (14:07:20 $-48$:20:36). FPOSS maximises the number of
observed stars, offering to the user the possibility to set different
priorities to the targets. To the input catalogue we attached a
priority list based on information obtained from the literature. We
tried to give highest priority to stars of spectral type earlier than late
G5, and they were deemed as probable cluster members. Cluster membership was 
originally assessed using the works by \citet{tycho} \citet{kharchenko2004}, 
\citet{kharchenko2005}, and \citet{dias2006}, all based mainly on the star's 
apparent proper motions. UVES fibers were positioned on four stars that, 
from previous literature \citep[e.g.][]{john2006}, we suspected to be 
chemically peculiar with low \vsini\ values. The remaining four UVES 
fibers were placed to obtain sky spectra, as requested by FPOSS.
We note that the number of MEDUSA fibers was larger than the
number of targets in our list of probable members, therefore we decided to 
place all remaining free fibers on targets more or less randomly selected 
in the instrument field of view. For this reason the final number of observed 
targets was larger than our list of probable members

After the observations were obtained, we performed a preliminary spectral 
analysis of all observed targets, and derived an estimate of the effective 
temperature, \vsini, and radial velocities (see Sect.~\ref{abn analysis}) from
a quick analysis of the H$\beta$ line observed with the GIRAFFE LR03 
setting. We then considered all targets with \Teff$>$5400\,K and radial 
velocity between $-35$ and 5\,\kms, and which were included in our original 
higher priority list because, based on proper motion criteria, were deemed 
probable cluster members. The radial velocity range [$-35$,$-$5]\,\kms\ was 
chosen based on the previously published value of the mean cluster radial 
velocity \citep[$-12.70\pm2.18$\,\kms][]{kharchenko2005} and on our 
preliminary \vr\ measurements of the observed brightest stars (considered 
fiducial cluster members). Out of 104 observed targets, we were left with a 
list of 38 stars.

Finally, we cross-checked our target list with the list of probable cluster
members compiled by \citet{claria93} on the basis of photometric
measurements, and we removed another 14 stars from our list. In this process 
we considered as non cluster members also the stars not listed in Claria's 
work. Table~\ref{Tab_Observations} lists the targets that have survived our 
selection criteria.

In this paper we report the spectral analysis performed on the 24
targets of Table~\ref{Tab_Observations}, which are cluster members
according to considerations based on proper motion, radial velocity,
and photometry, and have \Teff$>$5400\,K. We note that some
early-type cluster members may have been missed while preparing the
observations (either by accident, or because outside of the instrument
field of view), and our selection based on radial velocity has left
out some binaries that are potentially cluster members. Therefore,
Table~\ref{Tab_Observations} cannot be considered a complete list of
cluster members with spectral type earlier than G5.
\section{Abundance analysis}\label{abn analysis}
Model atmosphere calculations were performed with the \llm\ stellar model
atmosphere code \citep{llm}, assuming Local Thermodynamical Equilibrium 
(LTE), plane-parallel geometry, and modeling convection theory according 
to the \citet{cm1,cm2} model of convection 
\citep[see][for more details]{heiter}. We used the \vald\ database 
\citep{vald1,vald2,vald3} as a source of atomic line parameters for 
opacity calculations.

We derived the fundamental parameters for each analysed star mainly by fitting
synthetic line profiles, calculated with \synth\ \citep{synth3}, to the 
observed hydrogen lines: H$\beta$ and H$\gamma$ for the stars observed
with FLAMES/UVES, and H$\beta$ for the stars observed with FLAMES/GIRAFFE.
FLAMES/UVES spectra cover each the two hydrogen lines on more than one 
order, therefore we merged the unnormalised orders before the normalisation, 
which was performed with a low degree polynomial function. The H$\beta$ lines, 
observed with FLAMES/GIRAFFE, were normalised with a low degree polynomial 
function (usually of degree 4) to carefully selected continuum points. Our 
hydrogen line normalisation was performed on the basis of continuum 
points which extend for several Angstroms both blueward and redward of the 
hydrogen line wings. This ensures a high quality continuum normalisation of 
the hydrogen lines, thanks also to the absence of instrumental artifacts, 
such as scattered light, which would strongly reduce the precision of the 
normalisation.

\begin{figure}
%\sidecaption
\begin{center}
\includegraphics[width=80mm,clip]{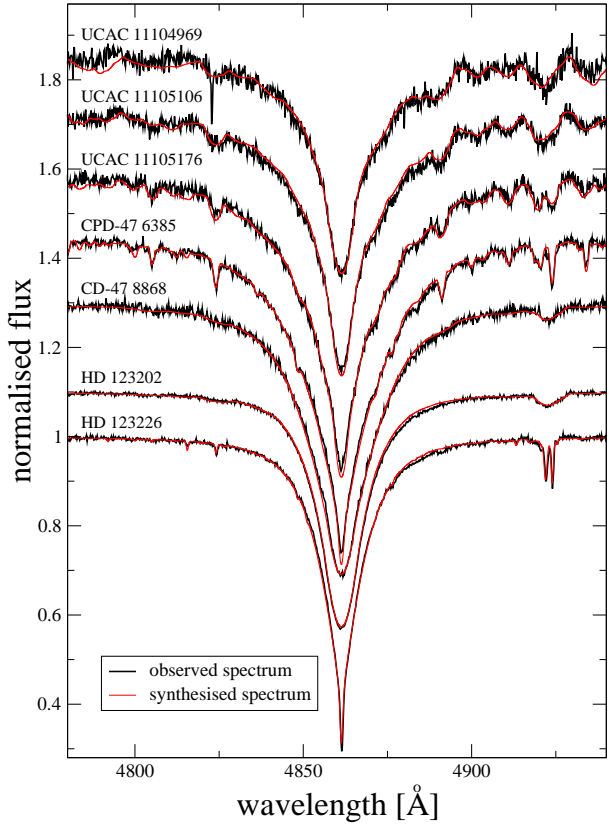}
\caption{Comparison between the observed H$\beta$ line profile (thick  
lines) and synthetic line profile (thin lines) for seven stars observed with
FLAMES/GIRAFFE. From top to bottom the stars here displayed are: UCAC~11104969
(\Teff\ = 7050\,K), UCAC~11105106 (\Teff\ = 7500\,K), UCAC~11105176 
(\Teff\ = 8100\,K), CPD-47~6385 (\Teff\ = 9350\,K), CD-47~8868 
(\Teff\ = 10400\,K), HD~123202 (\Teff\ = 11400\,K) and HD~123226 
(\Teff\ = 12400\,K). Profiles are vertically offset for display purpose.} 
\label{hbeta} 
\end{center} 
\end{figure}

Figure~\ref{hbeta} shows a comparison between observed and synthetic H$\beta$
line profiles for seven stars of various \Teff\ observed with FLAMES/GIRAFFE.
%--------------------------------------------------------------------
%--------------------------------------------------------------------

Due to the very broad temperature range covered by the analysed stars, the
hydrogen lines were used as both temperature and gravity indicators, because 
at lower temperatures hydrogen lines are more sensitive to temperature
variations, while at higher temperatures gravity has the larger influence. 
This is also reflected on our error bars on the fundamental parameters.

Since with the fitting of the hydrogen line wings it was not always possible 
to derive simultaneously both \Teff\ and \logg, we adopted also two other 
spectroscopic indicators based on the analysis of metallic lines. In 
particular, \Teff\ was determined eliminating the correlation between line 
abundance and the line excitation potential (\exc) for a given ion/element 
and the surface gravity was determined with the ionisation balance of 
several elements. In particular we adopted mainly Fe lines to 
derive both \Teff\ and \logg\ from metallic lines, adopting then Ti, Ca
and Si lines (the latter for the hotter stars of the sample) as secondary
indicators. Given the fact that the number of measurable lines decreases with 
increasing \vsini, for stars with a high projected rotational velocity we 
ended up adopting only Fe lines. 

Non-LTE effects might play an important role when measuring \logg\ by 
imposing ionisation equilibrium. \citet{fossati09} concluded that for solar
metallicity early-type stars \nlte\ effects for Fe are most likely to be 
negligible. \citet{mashonkina} performed \nlte\ calculations for Fe adopting 
the most complete iron model atom currently existing, and concluding that for
A-type main-sequence stars \nlte\ effects for FeI are less than 0.1\,dex and
negligible for FeII. Both Ti, Ca and Si are subject to \nlte\ effects, but we 
considered these elements only as secondary parameter indicator, giving 
most of the weight to what derived from Fe lines.

For the cooler stars we were not able to apply the method based on line 
profile fitting of gravity-sensitive metal lines with developed wings, 
usually adopted to derive \logg\ for late-type stars 
\citep[see e.g.][]{fuhrmann}, because the low resolution of the FLAMES/GIRAFFE 
spectra caused too heavy blending of those lines.

For stars with a low \vsini\  (\vsini$<30$\,\kms) LTE abundance 
analysis was based on equivalent widths, analysed with a modified version 
\citep{vadim} of the WIDTH9 code \citep{kurucz1993}, while for fast rotating 
stars we derived the LTE abundances by line profile fitting of carefully 
selected lines \citep[see][for more details]{fossati2007}.

The microturbulence velocity (\vmic) was determined by minimising the
correlation between equivalent width and abundance for several ions, for the
slowly rotating stars for which it was possible to measure the line equivalent 
widths, or following the procedure described in \citet{fossati2008} for fast 
rotators, except for HD~123202, for which we were not able to measure \vmic,
which we then assumed to be 0.0\,\kms, with an uncertainty of 1.0\,\kms.
%--------------------------------------------------------------------
\input{./Table_Parameters.tex}
%--------------------------------------------------------------------

Both the projected rotational velocity (\vsini) and \vr\ were determined by 
fitting synthetic spectra of several carefully selected lines on the observed 
spectrum. 

Table~\ref{Tab_Parameters} lists the fundamental parameters derived for 
each analysed star in NGC\,5460. For some stars a preliminary spectral 
classification is given here for the first time on the basis of the derived 
\Teff. The uncertainties of \Teff\ and \logg, listed in 
Table~\ref{Tab_Parameters}, were determined from the hydrogen line fitting, 
our main source for the parameter determination, therefore the given error 
bars take into account the quality of the observed spectra and the sensitivity
of the hydrogen line wings to both \Teff\ and \logg\ at different temperatures. 
Table~\ref{Tab_Parameters} also shows that the uncertainties on \Teff\ and 
\logg\ depend on the stellar \vsini. This is due to the fact that we also 
adopted metallic lines to determine the atmospheric parameters and the 
precision with which we determined the element/ion abundances decreases
with increasing \vsini, for example because of the decreasing number of
measurable lines for each ion.

The abundances obtained for each analysed star are shown in
Table~\ref{ngc5460 abn memb}. The error bars are the standard
deviations of the mean abundance derived from the individual line
abundances, which include also the error bars in oscillator strengths
determination given for the laboratory data \citep{fossati09}. 

The standard deviation of the mean underestimates the actual abundance 
error bars, since uncertainties on the atmospheric parameters should 
also be taken into account. In particular for the fast rotating stars, the
uncertainty on \vmic\ has a large impact on the total abundance uncertainty.
For two stars with very similar atmospheric parameters, but different \vsini\ 
values (CPD-47\,6385 and CD-47\,8905) we determined the element abundances
adopting model atmospheres for which we increased each atmospheric parameter 
one at a time, by an amount representative of the whole sample of stars: 
250\,K for \Teff, 0.2 for \logg\ and 0.7\,\kms\ for \vmic. 
Table~\ref{Tab_Errors} shows the results of this test.
For the slow rotator, the abundance error bar is highly dominated by the
uncertainty on \Teff, while for the fast rotator, the uncertainty on \vmic\ 
plays a crucial role. In both cases the uncertainty on \logg\ is almost
negligible. This test also shows the dependency of the abundance error bars 
on the lines selected to perform the abundance analysis: for the slow rotator
the abundance error bars of Mg, Ti and Ba, due to the uncertainty on \vmic, 
are rather large. This is due to the fact that most of the lines selected to 
measure the abundance of these three elements are strong (e.g. the Mg 
triplet at $\lambda\lambda\sim$\,5167, 5172 and 5183), being therefore very 
sensitive to \vmic\ variations. Anyway, this effect is mitigated by the fact 
that for the slow rotators the uncertainty on \vmic\ is not as large as 
0.7\,\kms, value chosen here to perform this test. Using the propagation 
theory in Table~\ref{Tab_Errors} we considered the situation where the 
determination of each fundamental parameter is an independent process, which 
in reality is not true, since there is a level of correlation between the 
various parameters. This makes the global uncertainties, given in columns 
five and nine of Table~\ref{Tab_Errors}, upper limits. In conclusion, 
following the results of this test and of the ones performed by 
\citet{fossati2008} and \citet{fossati09}, we can safely set a mean abundance 
uncertainty of 0.2\,dex, for the slow rotators, which increases linearly with 
\vsini, due to the increasing importance of the \vmic\ uncertainty in the 
total abundance error bar budget. The mean abundance error bars, adopted for 
each star, are shown in the last column of Table~\ref{ngc5460 abn memb}.

The impact of systematic uncertainties is different for different elements, 
but it is out of the scope of this work to go in such details, since here we 
are mainly interested in general trends and only marginally in the detailed 
abundance characteristics of each single star, which should require spectra 
with much higher resolution and signal-to-noise ratio (SNR), as demonstrated 
by \citet{fossati09}.
%--------------------------------------------------------------------
\input{./Table_Abundances.tex}
\input{./Table_Errors.tex}
%--------------------------------------------------------------------
%
\section{The impact of systematic errors}\label{comparison}
The scattering of the element abundance obtained from different lines
provides only a lower limit for the actual error in the abundance
determination. Systematic effects introduced by the normalisation to
the continuum, and inaccurate \Teff\ and \logg\ estimates may have a
large impact. To explore the reliability of the error estimates of
Table~\ref{ngc5460 abn memb} we re-analysed a number of stars in a
completely independent way, using the {\sc Zeeman}
\citep{john1988,wade2001} spectrum synthesis code. This `secondary'
analysis helps to ensure our results and estimated uncertainties are robust.

The {\sc Zeeman} code performs polarised radiative transfer, under the
assumption of LTE, using plane-parallel model atmospheres. Input LTE
model atmospheres were calculated using {\sc atlas9}
\citep{kurucz1993}, assuming solar abundances, which are a reasonable
starting point as none of the stars in NGC\,5460 show extreme chemical
peculiarities. Input atomic data was drawn from the \vald\ database,
using an `extract stellar' request, with effective temperature and
surface gravity tailored to the specific star, and solar abundances.

Initial values for temperatures and surface gravities were found by
fitting synthetic Balmer lines to the observations. These values were
confirmed, and when possible refined, by fitting metallic lines with a
range of ionisation states and excitation potentials. The Balmer line
fitting was performed by eye, though the metallic line fitting was
done using $\chi^2$ minimisation.

Abundances, \vsini\ and \vmic\ were determined by fitting a synthetic 
spectrum to the observation, using a Levenberg-Marquardt $\chi^2$ 
minimisation procedure. This was performed for 3 segments of spectrum 
between 100\,\AA\ and 500\,\AA\ long in the 4500--6000\,\AA\ region.  
The final best fit values reported here are averages over the individual 
segments. Uncertainties are generally taken to be the standard deviation.   
For elements with less then $\sim3$ useful lines, an uncertainty was 
estimated by eye, taking into account the scatter between lines, blending, 
and potential normalisation errors. 

\begin{figure}
\begin{center}
\includegraphics[width=80mm,clip]{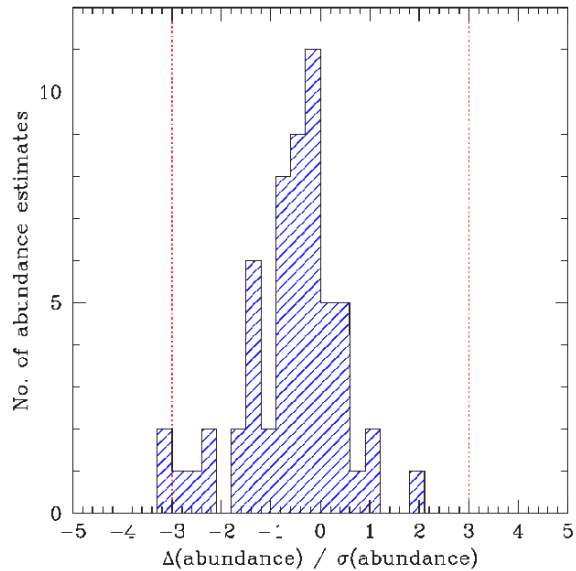}
\caption{Histogram of the difference of the abundances obtained with the 
secondary and the standard analysis, divided by the error bar of the standard 
analysis, listed in the last column of Table~\ref{ngc5460 abn memb}. The 
dashed lines correspond to a 3-$\sigma$ difference.}  
\label{comparison-abn-plot} 
\end{center} 
\end{figure}

Best fit results for the stars analysed with this method are presented in 
Table~\ref{comparison-abn-tab}. Figure~\ref{comparison-abn-plot} shows the
histogram of the difference of the abundances obtained with the secondary 
and the standard analysis, divided by the error bar of the standard 
analysis (last column of Table~\ref{ngc5460 abn memb}). Generally the results
are very consistent, with the abundances being within uncertainty of 
each other, at least within a $1.5\sigma$ limit. 

This secondary analysis supports also the conclusion that HD~123182 is
chemically peculiar. A large Mn overabundance is confirmed, though no
Sc or Hg abundances could be derived. Our secondary analysis confirms
also that HD~123183 is a chemically normal star, with a very high
\vsini, despite one published marginal magnetic field detection. In
both HD~123201B and HD~123184 discrepancies in \vsini\ were again
detectable for lines with high excitation potentials, particularly He
lines. The secondary analysis confirms both these stars are chemically
normal.
%--------------------------------------------------------------------
\input{./Table_Second_ABN.tex}
%--------------------------------------------------------------------
%--------------------------------------------------------------------
%--------------------------------------------------------------------
%
%
\section{Results}
Our \vr\ measurements allowed us to derive the mean cluster radial
velocity: $-17.9\pm5.2$\,\kms, obtained without taking into account
the double line spectroscopic binaries (SB2) stars. This value agrees rather 
well with $-12.70\pm2.18$\,\kms, published by \citet{kharchenko2005}, in 
particular taking into account that the latter value is the average of the 
\vr\ of only three stars.

We cannot exclude that some of the analysed stars are single line 
spectroscopic binaries (SB1), but this was impossible to establish with 
the available spectra. 

In the following we describe more in detail the results obtained for
the most interesting objects.
\subsection{HD~122983}
The star HD~122983 was observed by \citet{bagnulo2006} to measure the 
longitudinal magnetic field, but no field was detected. The abundance pattern 
shows that HD~122983 is a He-weak star, but it was not possible to determine 
the kind of chemical peculiarity (e.g. HgMn star). Manganese lines are too 
shallow to be measured and none of the covered mercury line is visible. 
A spectrum covering the spectral region around 3900\,\AA\ would be extremely 
valuable to detect the presence of Hg overabundance \citep{hubrig1996}. 
We looked also for the overabundance of elements characteristic of HgMn stars, 
such as P and Xe, but without results. 
\subsection{HD~123182}
The abundance pattern of HD~123182 reveals that the star is most likely 
a mild He-weak star, in particular it could be a hot Am star or a cool 
HgMn star. Sc lines are too weak to be measurable, so it is not possible 
to say whether it is underabundant or not. The not extreme overabundance 
of Cr and Fe (as expected for the rather high \vsini), and the fact 
that Hg, P and Xe are not visible (although lines would be covered) are 
in agreement with an Am classification. On the other hand the clear Mn 
overabundance would tend more to a HgMn classification. On the basis of 
the available data we can only conclude that HD~123182 is a He-weak star 
of an unclear type. This star could also be a single line spectroscopic 
binary, due to the high radial velocity, compared to the one of the 
other cluster stars.
\subsection{HD~123183}
The star HD~123183 was observed with FORS1 by \citet{bagnulo2006} who 
published a marginal magnetic field detection of 
$\langle Bz\rangle=-440\pm146$\,G. The abundance pattern does not reveal any 
chemical peculiarity that could strengthen this detection. The very high 
\vsini\ of $275\pm14$\,\kms\ also goes against a classification of this 
object as chemically peculiar. For this reason we conclude that HD~123183 
is a chemically `normal' early-type star. Given the very high \vsini, it 
would be very valuable to perform new magnetic field measurements.
\subsection{HD~123225}
\begin{figure}
\begin{center}
\includegraphics[width=80mm,clip]{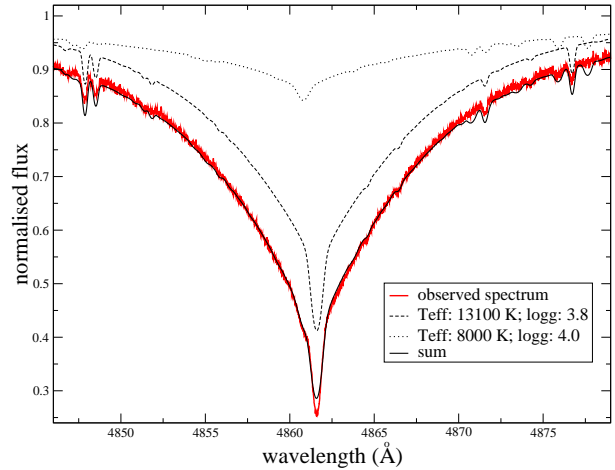}
\caption{Fit of the observed H$\beta$ line profile for HD~123225. The red
line shows the observed spectrum, while the dashed and dotted lines are the
synthetic spectra of the two components assuming \Teff\ and \logg\ of 13100\,K
and 3.8 for the hotter star and 8000\,K ad 4.0 for the cooler star. The black 
solid line shows the sum of the two synthetic spectra.} 
\label{hbeta_hd123225} 
\end{center} 
\end{figure}

The FLAMES/UVES observations of HD~123225 revealed that this is a SB2 
system, where both components appear to have a rather low \vsini. The spectral
lines of the two components are well separated: \vr\ of the first component is
$\sim-$2\,\kms, while \vr\ of the second component is $\sim-53$\,\kms. With 
only one available spectrum it was not possible to perform a precise 
abundance analysis, but the large separation of the two components allowed to 
roughly estimate their \Teff: 13100\,K for the primary and 
8000\,K for the secondary star. These values were obtained from a fit of 
synthetic spectra on the observed H$\beta$ profile (see 
Fig.~\ref{hbeta_hd123225}), taking into account the expected relative 
difference in flux due to the different stellar radii of the two components and 
assuming that both are main sequence stars (\logg\ was set to 3.8 for the 
hotter star and to 4.0 for the cooler star).
%--------------------------------------------------------------------
%--------------------------------------------------------------------

The \vsini\ value of the primary star is $\sim$11\,\kms, while for the 
secondary star we derived a \vsini\ value of $\sim$20\,\kms. Given the 
roughly determined \Teff\ of the two components and their \vsini, it is 
possible that both stars developed chemical peculiarities and in particular 
HgMn peculiarities for the primary star and Am peculiarities for the secondary 
star. To reveal whether HgMn peculiarities are present in the primary star, 
we looked for typical signatures present in HgMn stars. The spectrum 
reveals strong Mn lines in addition to evidences of Hg lines, allowing to 
conclude that the primary component is a HgMn star. This conclusion is 
strengthened by the fact that we found several strong lines of both Ne, P 
and Xe, typical of HgMn stars. For the secondary star it was not possible 
to determine whether it shows Am chemical peculiarities since the Sc lines 
are too shallow to be visible. In any case, the fact that the star seems 
to have Fe lines compatible with a solar iron abundance, let us believe 
that the secondary component is not an Am star. 

HD~123225 was observed with FORS1 by \citet{bagnulo2006} and no magnetic 
field was detected.
\subsection{HD~123201B and HD~123184}
The stars HD~123201B and HD~123184 are both chemically `normal' early-type 
stars and share the common property of showing a different \vsini\ for the high
\exc\ lines, compared to the other spectral lines. The \vsini\ of HD~123201B, 
measured on low \exc\ lines, is $\sim$202\,\kms, while He lines are best fit 
with a \vsini\ of $\sim$150\,\kms. The same occurs for HD~123184, for 
which \vsini\ is $\sim$60\,\kms, but He lines are best fit with a 
\vsini\ of $\sim$50\,\kms. 

\begin{figure}
\begin{center}
\includegraphics[width=80mm,clip]{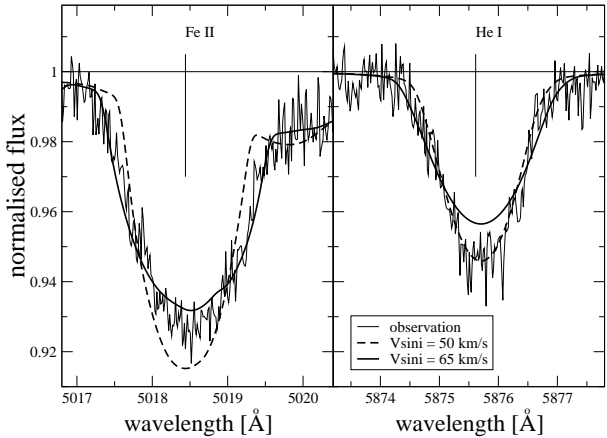}
\caption{Comparison between the observed spectrum (thin line) and synthetic 
spectra calculated with \vsini\ = 60\,\kms\ (thick line) and 
\vsini\ = 50\,\kms\ (thick dashed line) in the region of the FeII 
line at $\sim$5018\,\AA\  (left plot) and of the HeI multiplet at 
$\sim$5875\,\AA\  (right plot), for HD~123184.} 
\label{He-vsini} 
\end{center} 
\end{figure}

Figure~\ref{He-vsini} shows the comparison between the observed spectrum 
and two synthetic line profiles with \vsini\ = 60 and 50\,\kms\ in 
the region of the FeII line at $\sim$5018\,\AA\ and of the HeI 
multiplet at $\sim$ 5875\,\AA\ for HD~123184. The plot shows that for the
HeI multiplet the adopted stellar \vsini\ of 60\,\kms is far too high, 
while a \vsini\ of 50\,\kms\ is necessary to fit the observed line profile. 
In contrast, the region around the FeII line shows that a \vsini\ of 
60\,\kms\  is required to fit this line. For both stars the He abundance was 
measured with the \vsini\ determined for the He lines.
%--------------------------------------------------------------------
%--------------------------------------------------------------------

This effect is most likely a consequence of the fact that both stars are 
fast rotators with a rather small inclination of the rotation axis relative to
the line of sight, leading to a temperature difference between the poles 
(hotter) and the equator (cooler) \citep{zeipel}.
\subsection{UCAC~11105038}
This is an A-type star with a \vsini\ of 85$\pm$5\,\kms\ and a relatively high
\vmic\ of about 3.4\,\kms. Both \vsini\ and \vmic\ could indicate a possible Am
chemical peculiarity of this star, in particular \vsini\ is below 90\,\kms\ 
\citep{charbonneau}, and \vmic\ is typical of Am stars 
\citep{john1998,fossati2008}. Both Ca and Sc abundances are below the solar 
values, as well as the Ba overabundance is indicative of Am chemical 
peculiarities, but the underabundance of the Fe-peak elements goes against 
an Am classification. 
\section{Discussion}\label{discussion}
\subsection{Abundance patterns in NGC\,5460: from F- to B-type stars}
\label{metallicity ngc5460}
\begin{figure}
\begin{center}
\includegraphics[width=80mm,clip]{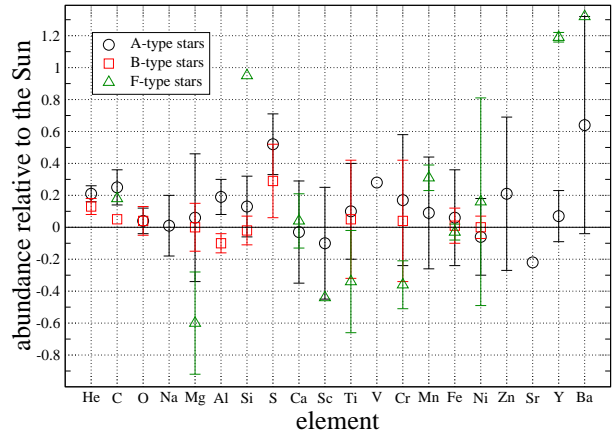}
\caption{Comparison between the mean abundances obtained for B-, A- and
F-type stars indicated by an open square, circle and triangle respectively. 
The error bars are the standard deviations from the calculated mean abundances. 
The abundance patter of the mean F-type stars is corresponds to the abundance
pattern of UCAC~11104969, since it is the only F-type stars of the sample with
measured abundances.} 
\label{mean_BAF} 
\end{center} 
\end{figure}

Figure~\ref{mean_BAF} shows the comparison among the mean abundance pattern
obtained for B-, A- and F-type stars here analysed and considered members of 
the cluster. The error bars are the standard deviation from the mean abundance. 
In the whole sample, after the assignment of the cluster membership and 
binarity, only one F-type single star, UCAC~11104969, was left in the sample, 
therefore the mean abundance of the F-type stars corresponds to the abundance 
pattern of UCAC~11104969.
%--------------------------------------------------------------------
%--------------------------------------------------------------------

Figure~\ref{mean_BAF} shows that the three abundance patterns are quite 
similar and concentrated around solar values except for a sulfur 
overabundance, which cannot be ascribed to \nlte\ effects \citep{fossati2007}, 
and for an apparent Am peculiarity of some elements in UCAC~11104969, excluded
by the high \vsini\ value (260$\pm$15\,\kms).

An important result of this work is the estimation of the cluster
metallicity ($Z$), spectroscopically never measured before. For the elements 
that most affect $Z$, such as C, O, Si, and Fe, we registered a rather good 
agreement among the three stellar types, with the exception of the Si 
abundance of UCAC~11104969\footnote{the star has a high \vsini\ and the 
Si abundance was derived only from one line, making the Si abundance value 
rather uncertain}, which was not taken into account in the determination 
of $Z$. Since the cluster metallicity should be calculated on the basis of 
the abundances of the cool stars (more mixed) and that only one F-type star 
is present in our sample, we decided to derive $Z$ from the mean abundance of 
UCAC~11104969 and of the A-type stars with \Teff\ $\leq$ 8500\,K.

The cluster metallicity ($Z$) is defined as follows:
\begin{equation}
\label{Z}
Z_{\rm cluster}=\frac{\sum_{a \geq 3}m_{a}10^{\log(N_{a}/N_{tot})}}{\sum_{a \geq 1}m_{a}10^{\log(N_{a}/N_{tot})}},
\end{equation}
where $a$ is the atomic number of an element with atomic mass m$_{\rm a}$.
Making use of the mean abundances, we derived a metallicity of 
$Z$ = 0.014 $\pm$ 0.003, consistent with the solar $Z$
\citep[$Z$=0.012,][]{met05}, adopting solar abundances by \citet{met05} 
for all elements that were not analysed. 

The $Z$ value adopted by several other authors and used to characterise 
isochrones is calculated with the following approximation:
\begin{equation}
\label{clusterZ}
Z_{\mathrm cluster} \simeq 10^{([Fe/H]_{\mathrm Fstars}-[Fe/H]_{\odot})} \cdot Z_{\odot}\ ,
\end{equation}
assuming $Z_{\odot}$=0.019, which is favored by the theoretical stellar 
structure models based on \citep{anders1989} solar abundances. 
We recalculated the $Z$ value of NGC\,5460, according to this approximation, 
obtaining $Z$ = 0.013$^{+0.008}_{-0.005}$, only about consistent with a solar 
metallicity. This value was derived from the mean Fe abundance of the seven 
stars taken into account ([Fe/H] = $-0.18\pm$0.22\,dex). We believe that a 
solar metallicity would be anyway more appropriate for NGC\,5460. The Fe 
underabundance of $-$0.18\,dex is biased by two stars (out of a sample of 
seven) which show a considerable underabundance ([Fe/H]$\sim-$0.46\,dex). On 
the other hand, for the other five stars the iron abundance is 
[Fe/H]=$-$0.06$\pm$0.12\,dex, which leads to $Z$ = 0.017$^{+0.005}_{-0.004}$, 
that, given the uncertainties, is perfectly consistent with a solar 
metallicity. In addition, Fig.~\ref{mean_BAF} shows that the mean Fe 
abundance of the analysed cluster stars is well centered around the solar 
value.
\subsection{HR diagram}\label{hr diagram}
\begin{figure}
\begin{center}
\includegraphics[width=80mm,clip]{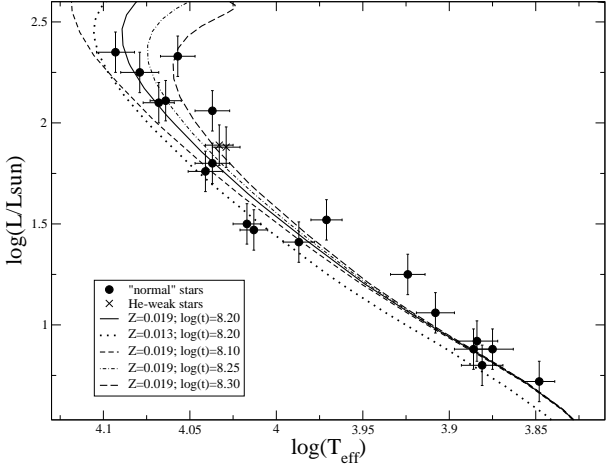}
\caption{Hertzsprung-Russell diagram of NGC\,5460. The crosses 
and filled circles show the position on the HR diagram for the chemically
peculiar and chemically normal stars, respectively. The error bar in 
luminosity is 0.10 dex. The full and dotted lines show isochrones from 
\citet{marigo} for an age of \logt\ = 8.20 with solar metallicity 
($Z$ = 0.019) and $Z$ = 0.013, respectively. The short dashed, thin dot-dashed
line and long dashed line show solar metallicity isochrones of three different
ages: \logt\ = 8.10, 8.25 and 8.30, respectively.} 
\label{hr_diagram-ngc5460} 
\end{center} 
\end{figure}

With the obtained effective temperatures and the magnitudes provided by 
\citet{claria93}, we built the Hertzsprung-Russell (HR) diagram of the 
cluster (Fig.~\ref{hr_diagram-ngc5460}), adopting a distance modulus of 
9.44\,mag, a reddening of 0.092\,mag 
\citep[both values given by WEBDA,][]{webda}, and the bolometric correction 
from \citet{balona}.
%--------------------------------------------------------------------
%--------------------------------------------------------------------

Luminosities are listed in Table~\ref{table luminosity-ngc5460}. With a 
distance modulus uncertainty of 0.20\,mag, an uncertainty in the bolometric 
correction of about 0.07\,mag, and a reddening uncertainty of 0.01\,mag, we 
estimate the typical uncertainty in \mbol\ being about 0.28\,mag, 
corresponding to an uncertainty in \logl\ of about 0.10\,dex.
%--------------------------------------------------------------------
\input{./Table_Luminosities.tex}
Two stars of our sample, HD~122983 and HD~123183, are included in 
the work by \citet{john2006} and there is a good agreement between the 
obtained \logl\ and \M\ values, where the small differences are due to the 
adoption of a different source for the stellar magnitudes. We adopted the 
magnitudes given by \citet{claria93}, while \citet{john2006} adopted the 
magnitudes published by \citet{claria1971}.

Figure~\ref{hr_diagram-ngc5460} shows that assuming a metallicity of 0.013 
would have little impact on the estimate of the cluster age, but just of 
the cluster distance, decreasing it by about 100\,pc. Therefore, we can 
safely conclude that the distance of the cluster is within the range 
670--770\,pc. Out HR diagram allows also to better constrain the cluster age.
The age of the solar metallicity isochrone fitting the cluster main sequence 
is \logt=8.10, which we can then consider as the minimum cluster age. The 
maximum age instead depends strongly on which stars of the upper main 
sequence are blue strugglers. If none of our analysed cluster stars is a blue 
struggler, the maximum cluster age is most likely to be \logt=8.25, while if 
HD\,123226 (the hottest star in our sample) is a blue struggler the maximum 
cluster age is more likely to be \logt=8.30. On the basis of these 
considerations, our best estimate of the cluster age is \logt=8.20$\pm$0.10, 
in comparison to \logt=8.20$\pm$0.20 previously given by \citet{ahumada2007}.
\subsection{Abundances vs. \Teff\ and \vsini}\label{correlation parameters}
\begin{figure}
\begin{center}
\includegraphics[width=80mm,clip]{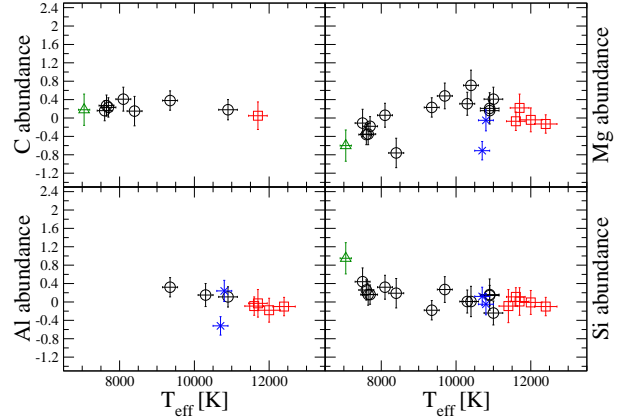}
\caption{Abundances relative to the Sun \citep{met05} of C, Mg, Al 
and Si as a function of \Teff\ for B-type stars (open squares), A-type stars 
(open circles), F-type stars (open triangles) and He-weak stars (crosses).} 
\label{trends-teff-ngc5460-1} 
\end{center} 
\end{figure}

\begin{figure}
\begin{center}
\includegraphics[width=80mm,clip]{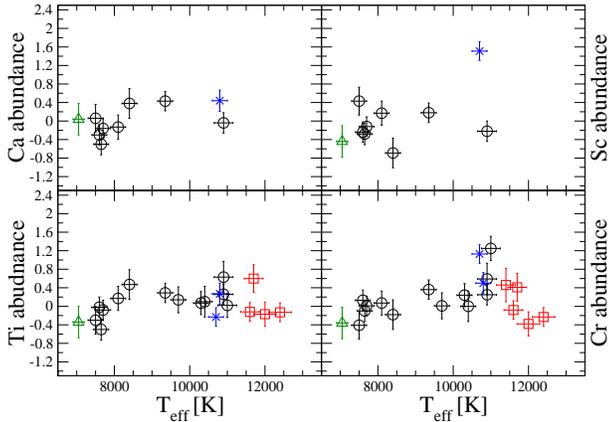}
\caption{Same as Fig.~\ref{trends-teff-ngc5460-1}, but for Ca, Sc, Ti and Cr.} 
\label{trends-teff-ngc5460-2} 
\end{center} 
\end{figure}

\begin{figure}
\begin{center}
\includegraphics[width=80mm,clip]{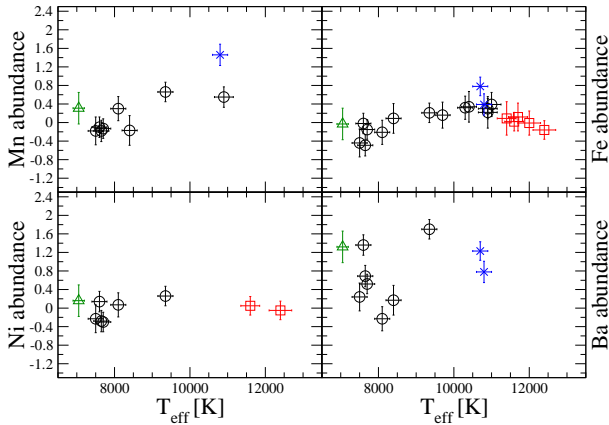}
\caption{Same as Fig.~\ref{trends-teff-ngc5460-1}, but for Mn, Fe, Ni and Ba.} 
\label{trends-teff-ngc5460-3} 
\end{center} 
\end{figure}

From Fig.~\ref{trends-teff-ngc5460-1} to Fig.~\ref{trends-teff-ngc5460-3} we 
display the abundances of chemical elements versus \Teff.
%--------------------------------------------------------------------
%--------------------------------------------------------------------
%--------------------------------------------------------------------
%--------------------------------------------------------------------
%--------------------------------------------------------------------
%--------------------------------------------------------------------

For some elements, such as Mg and Fe, it is possible to notice a correlation 
of the abundances with \Teff, with an abundance increase with \Teff\ up to 
\Teff$\sim$10500\,K and an abundance decrease for higher temperatures. A 
linear fit shows that these correlations are significant for both Mg and Fe 
(see Table~\ref{correlations}).
%--------------------------------------------------------------------
\input{./Table_Correlations.tex}
%--------------------------------------------------------------------

\begin{figure}
\begin{center}
\includegraphics[width=80mm,clip]{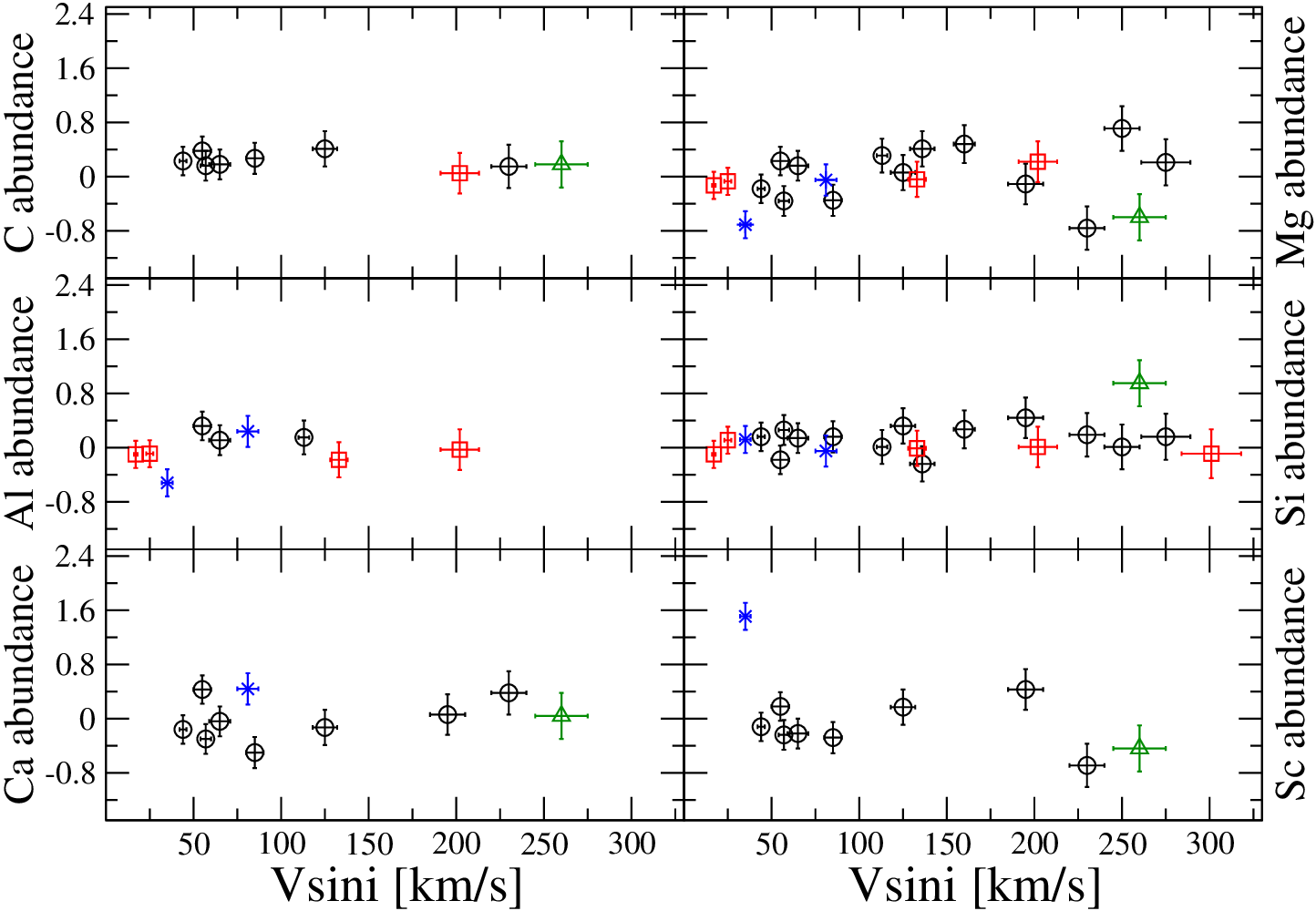}
\includegraphics[width=80mm,clip]{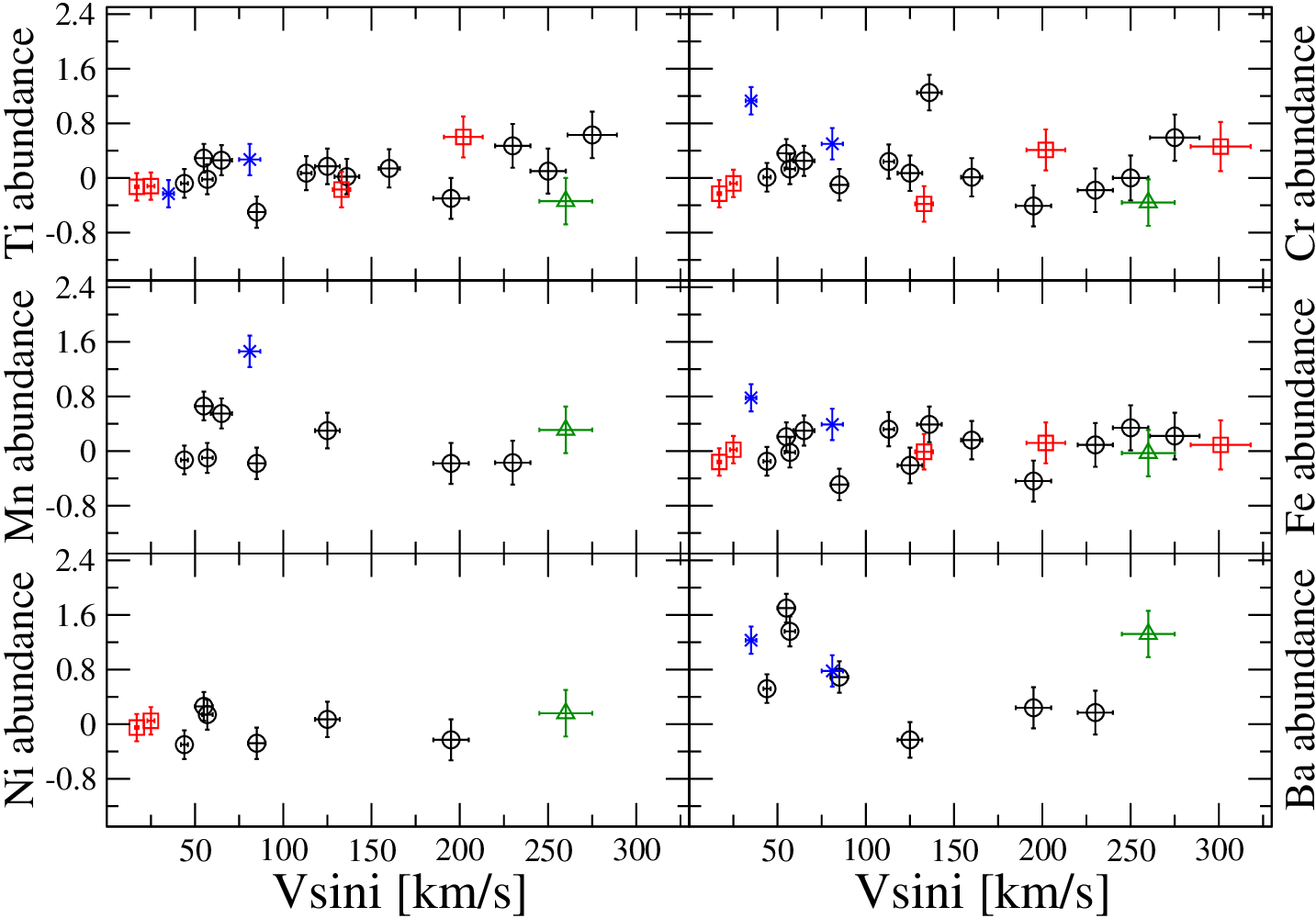}
\caption{Abundances relative to the Sun \citep{met05} as a 
function of \vsini\ for B-type stars (open squares), A-type stars 
(open circles), F-type stars (open triangles) and He-weak stars (crosses).} 
\label{trends-vsini-ngc5460} 
\end{center} 
\end{figure}

These correlations could in principle not be real, but due to other factors, 
such as \vmic, \logg\ and \vsini. We excluded that both \vmic\ and \logg\ 
produced such trends, since the abundance uncertainties take into account 
the \vmic\ uncertainty, and the effect of \logg\ variations on the abundances
is much smaller than the deviations obtained here. In addition, we did not 
find any clear correlation between the abundances and \vsini, as shown in 
Fig.~\ref{trends-vsini-ngc5460}, except for a decreasing Ba abundance 
with increasing \vsini\ up to \vsini\ values of about 125\,\kms, in agreement 
with predictions by \citet{TC93}. We find also unlikely that our LTE 
approximation could be the source of the abundance correlations with \Teff,
since for both Mg and Fe the expected \nlte\ corrections are small
\citep{przybilla01,fossati09} and their dependence on \Teff\ is not expected 
to be as pronounced as shown by our results. 
%--------------------------------------------------------------------
%--------------------------------------------------------------------

\begin{figure}
\begin{center}
\includegraphics[width=80mm,clip]{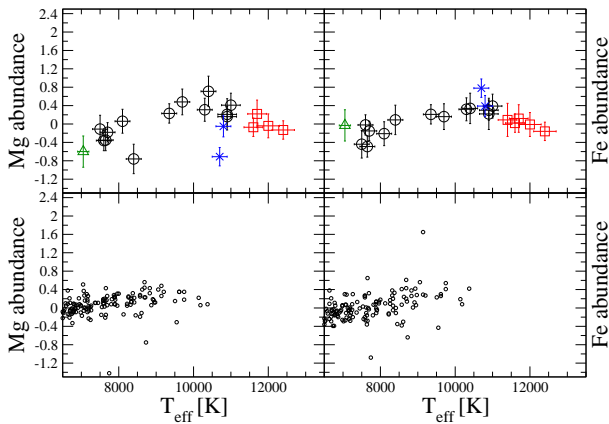}
\caption{Comparison between the results obtained in this work (upper panels) 
and by \citet{erspamer} (lower panels), for Mg (on the left side) and Fe (on 
the right side). B-type stars are denoted with open squares, A-type stars with 
open circles, F-type stars with open triangles and He-weak stars with crosses.} 
\label{comp-erspamer} 
\end{center} 
\end{figure}

As further check for the obtained abundance correlations with \Teff, we 
looked for similar trends in other consistently analysed sample of stars 
covering a large temperature range. 
\citet{erspamer} analysed the spectra of 140 A- and F-type stars, present in 
the ELODIE archive\footnote{http://atlas.obs-hp.fr/elodie/}, obtaining for each
star fundamental parameters and abundances of several elements. 
Figure~\ref{comp-erspamer} shows a comparison between the results we obtained 
in NGC\,5460 and what published by \citet{erspamer} for Mg and Fe.
%--------------------------------------------------------------------
%--------------------------------------------------------------------

We register an agreement for both elements, in particular for Fe (see 
Table~\ref{correlations}), although the trends appear less pronounced in the 
results published by \citet{erspamer}. Unfortunately the sample of stars 
analysed by \citet{erspamer} did not cover stars with temperatures higher 
than 10500\,K, so that it was not possible to check the presence of the 
abundance drop, present in our results.

Diffusion processes could be responsible for the obtained abundance 
correlations with \Teff. With the increase of the effective temperature 
the hydrogen and helium ionisation zones shift more and more towards the 
stellar surface, and as a consequence they become thinner and less turbulent. 
In this way diffusion processes become more and more effective. For example, 
it is well known that Ba is overabundant in early-type stars, which is a clear 
signature of an influence of diffusion even in chemically normal stars 
\citep{fossati09}. Another example is the presence of a correlation 
between strontium and oxygen or strontium and magnesium, obtained for 
chemically normal B-type stars by \citet{HH2003} (strontium is radiatively 
driven outwards, Mg and O sink down).

The efficiency of diffusion processes is reduced by rotation. In
particular, simulations carried out by \citet{TC93} show that for rotational
velocities higher than 125\,\kms\ the envelope is completely mixed.
This would suggest that correlations between abundances of diffusion indicator
elements (such as Sr, O, Mg, and Ba) and \vsini\ should be obtained even for 
chemically normal star. But correlations between abundance and \vsini\ are 
not found here, except for Ba, one of the diffusion indicators, for which 
the abundance decreases mildly with \vsini, up to \vsini\ values of 
$\sim$125\,\kms, in agreement with what predicted by \citet{TC93}. Probably 
a much larger sample of stars, which includes chemically normal low \vsini\ 
stars could show the presence or not of these correlations.
\section{Conclusions}\label{conclusions}
We obtained low and mid-resolution spectra of several stars in the field of 
the NGC\,5460 open cluster with the FLAMES spectrograph at the ESO/VLT, and
we selected 24 probable cluster members with \Teff$>7000$. This
cluster was never previously spectroscopically studied.

We calculated the fundamental parameters and performed a detailed 
abundance analysis for all the stars of the sample, except for three
objects (HD~123225, Cl NGC\,5460~BB\,338, and UCAC~11105152), which turned out 
to be double line spectroscopic binaries. Two stars, HD122983 and 
HD~123182, are He-weak stars, and we suspect that the hotter component of 
the HD~123225 SB2 system is a HgMn chemically peculiar star. 
\citet{bagnulo2006} obtained a marginal magnetic field detection for 
HD~123183, but the star has remarkably high \vsini\ for an Ap star, and 
no clear chemical peculiarities. 

The mean abundance pattern obtained for B-, A- and F-type stars
considered members of the cluster concentrate around solar values. We
determined the cluster metallicity from the cooler stars of our sample
(\Teff$<$8500\,K). The resultant metallicity is $Z = 0.014 \pm 0.003$. 
We rederived $Z$ also from the mean iron abundance 
([Fe/H]$ = -0.18\pm0.22$\,dex), adopting the approximation given in 
Eq.~\ref{clusterZ}, obtaining $Z = 0.013^{+0.008}_{-0.005}$. We believe anyway 
that this metallicity is underestimated, due to two stars of the sample which 
show a particularly low iron abundance, and, without taking into account
these two objects, the metallicity turned out to be almost solar
($Z$ = 0.017$^{+0.005}_{-0.004}$).

For each star of the sample we derived photometrically \logl\ and,
adopting the spectroscopic temperatures derived for the abundance analysis,
we built the HR diagram of the cluster. From a comparison
between the cluster HR diagram and isochrones by \citet{marigo} we obtained 
that the cluster distance lies in the range 670--770\,pc and that the cluster
age is \logt=$8.20\pm0.10$, where the maximum age depends anyway on which stars
of the upper main sequence are blue strugglers.

We plotted the abundances of the elements analysed in most of the stars 
of our sample as a function of \vsini\ and \Teff. We did not find any clear
abundance trend with respect to \vsini, except for barium, for which we register
a decrease of the abundance with increasing \vsini, up to a \vsini\ of about
125\,\kms. For some elements, particularly Mg and Fe, we found trends between 
abundance and \Teff; their abundances increase up to a \Teff\ 
value of about 10500\,K, then decrease with increasing \Teff. This
abundance trend is marginally present also in the sample of stars analysed by
\citet{erspamer}, but their results do not cover the hotter region, where 
we found the abundance decline. We suggest that these trends
might be signatures of diffusion in early-type stars.
\section*{Acknowledgments}
Based on observations made with ESO Telescopes at the Paranal Observatory 
under programme ID~079.D-0178. Astronomy research at the Open University is 
supported by an STFC rolling grant (L.F.). C.P. acknowledges the support of the
project P19503-N16 of the Austrian Science Fund (FWF). O.K. is a Royal Swedish 
Academy of Sciences Research Fellow supported by grants from the Knut and 
Alice Wallenberg Foundation and the Swedish Research Council. LF is deeply in
debt with Denis Shulyak for the calculation of the \llm\ stellar model 
atmospheres. We also acknowledge the use of cluster facilities at the
Institute for Astronomy of the University of Vienna. We thank the anonymous
referee for the many useful comments which improved the original manuscript.

\bsp

\label{lastpage}

\end{document}

%% file: Table_Observations.tex
\begin{table*}
%\begin{minipage}[t]{\columnwidth}
%\renewcommand{\footnoterule}{}  % to avoid a line before footnotes 
\caption[ ]{List of programme stars. The last three entries refer to SB2 stars for which no abundance analysis could be performed.
The $V$ magnitude is taken from \citet{claria93}. The symbol * in the spectral type column (No. 4) means that this was inferred from 
our determined effective temperature. SNR (col. 10) is the signal-to-noise ratio per spectral bin obtained with the 
GIRAFFE HR09/HR11/LR03 or the UVES 520\,nm settings.}
\label{Tab_Observations}
\begin{center}                      
%\rotatebox{90}{
\scriptsize{
\begin{tabular}{lrrccccclc}
\hline
\hline
            & Claria et   &   &     & Proper motion              & Membership  & Proper motion                     & Pkin/Pph/Psp &          &                  \\
            & al. (1993)  &   &Sp.  & Dias et al.               & Dias  et al.& Hog et al. (2000)                  & Kharchenko   &\ \ \ \ \ \vr  &               \\
STAR        & ident. \#   &$V$&Type & (2006) [mas/yr]           & (2006) [\%] & [mas/yr]                           & et al. (2004)&[\kms]    &  SNR               \\
\hline
HD~122983           & 142  &  9.87  &  B9 &$-5.6\pm0.9 ;           -0.8\pm1.0$&  60 &$-5.3\pm1.1 ;           -0.3\pm1.1$& 0.69/0.66/1 &$-24\pm2$ & 159/178/213  \\    
HD~123097           & 131  &  9.30  &  B9 &$-4.9\pm1.1 ;           -3.1\pm1.2$&  43 &$-4.6\pm1.2 ;           -2.4\pm1.1$& 0.90/1.00/1 &$-20\pm2$ & 209/233/275  \\    
HD~123182           &  87  &  9.88  &  B9 &$-5.8\pm1.1 ;           -3.2\pm1.2$&  44 &$-5.7\pm1.2 ;           -3.8\pm1.4$& 0.98/1.00/1 &$  0\pm4$ & 148/165/195  \\    
HD~123183           &  86  & 10.11  &  A0 &$-6.6\pm1.3 ; \phantom{-}0.8\pm1.3$&  58 &$-6.7\pm1.8 ;           -0.4\pm1.8$& 0.56/0.93/1 &$-19\pm7$ & 144 [UVES]   \\    
HD~123184           &  81  &  9.54  &  A0 &$-4.9\pm0.9 ; \phantom{-}2.4\pm0.9$&  51 &$-4.8\pm1.2 ;           -3.7\pm1.1$& 0.63/1.00/1 &$-15\pm3$ & 195 [UVES]   \\    
HD~123201B          &  85  &  9.55  &  B8 &$-3.6\pm1.3 ;           -5.4\pm1.3$&  20 &$-5.0\pm1.7 ;           -5.6\pm1.7$& 0.79/1.00/1 &$-20\pm5$ & 199 [UVES]   \\    
HD~123202           &  79  &  9.04  &  B9 &$-4.0\pm1.0 ;           -1.9\pm0.9$&  57 &$-4.2\pm1.1 ;           -2.7\pm1.1$& 0.68/1.00/1 &$-19\pm11$& 247/275/323  \\     
HD~123226           &  61  &  9.15  &  B8 &$-6.3\pm1.2 ;           -4.4\pm1.1$&  31 &$-5.9\pm1.6 ;           -6.3\pm1.5$& 0.66/1.00/1 &$-12\pm1$ & 231/258/306  \\    
HD~123269           &  31  &  9.55  &  B8 &$-7.1\pm1.0 ;           -2.8\pm1.2$&  44 &$-7.0\pm1.2 ;           -2.0\pm1.2$& 0.56/1.00/1 &$-20\pm2$ & 185/211/243  \\    
CD-47~8868	    & 141  & 10.66  &  A0 &$-6.4\pm1.4 ; \phantom{-}2.8\pm2.4$&  56 &$-6.3\pm1.7 ; \phantom{-}2.4\pm1.6$& 0.66/1.00/1 &$-10\pm9$ & 104/116/137  \\    
CD-47~8869	    & 136  & 10.94  &  A5*&$-4.9\pm1.7 ;           -0.6\pm1.4$&  60 &$-4.6\pm1.7 ; \phantom{-}0.4\pm1.7$& 0.89/1.00/1 &$-16\pm5$ & 96/106/125   \\    
CD-47~8889	    &  97  & 10.76  &  A0 &$-6.7\pm1.4 ;           -3.0\pm2.2$&  42 &$-7.9\pm1.7 ;           -5.1\pm1.7$& 0.64/0.67/1 &$-19\pm4$ & 99/110/128   \\    
CD-47~8905	    &  58  & 10.78  &  A0 &$-5.4\pm1.4 ;           -3.2\pm1.4$&  44 &$-5.2\pm2.1 ;           -3.1\pm2.0$& 0.97/0.64/1 &$-18\pm5$ & 100/111/131  \\    
CPD-47~6379         &  54  & 10.17  &  A0 &$  -                              $&   - &$       -                         $&   - /  - /- &$-24\pm7$ & 120/134/158  \\    
CPD-47~6385         &  59  & 10.46  &  A0 &$-5.3\pm1.6 ;           -2.5\pm1.4$&  52 &$-5.3\pm2.0 ;           -0.2\pm1.9$& 0.91/0.95/1 &$-19\pm2$ & 123/135/162  \\    
UCAC~11104969       & 133  & 12.01  &  F1*&$-6.6\pm2.1 ;           -1.4\pm1.7$&  57 &$-7.9\pm2.0 ;           -0.6\pm2.0$&   - /  - /- &$-17\pm3$ & 58/63/76     \\    
UCAC~11105038       & 125  & 11.59  &  A8*&$-5.3\pm1.9 ;           -0.2\pm1.4$&  60 &$-4.8\pm1.9 ; \phantom{-}1.3\pm1.9$&   - /  - /- &$-20\pm4$ & 69/78/94     \\    
UCAC~11105106       &  89  & 11.69  &  A8*&$-7.2\pm1.4 ; \phantom{-}0.4\pm1.4$&  52 &$11.1\pm2.2 ; \phantom{-}0.6\pm2.2$&   - /  - /- &$-16\pm3$ & 66/73/82     \\    
UCAC~11105176       &  91  & 11.33  &  A6*&$-7.4\pm1.4 ;           -1.0\pm2.7$&  50 &$-6.6\pm2.3 ; \phantom{-}1.1\pm2.2$&   - /  - /- &$-17\pm5$ & 78/87/104    \\    
UCAC~11105213       &  94  & 11.83  &  A8*&$-3.8\pm2.9 ;           -4.9\pm1.4$&  30 &$       -                         $&   - /  - /- &$-21\pm3$ & 62/69/79     \\    
UCAC~11105379       &  36  & 11.75  &  A8*&$-6.8\pm2.1 ; \phantom{-}2.4\pm1.4$&  56 &$-5.4\pm2.3 ; \phantom{-}2.1\pm2.1$&   - /  - /- &$-21\pm2$ & 65/72/87     \\  [2mm]  
\hline								
HD~123225           &  55  &  9.01  &  B9 &$-4.9\pm1.2 ;           -2.9\pm1.0$&  52 &$-6.0\pm1.2 ;           -2.4\pm1.3$& 0.99/1.00/1 &$-2\pm11\ / -53\pm2$& 207 [UVES] \\
Cl~NGC~5460~BB\,338 &  60  & 11.80  &  F* &  -                                &   - &    -                              &   - /  - /- &$-23\pm6\ / -23\pm1$&49/55/69    \\ 
UCAC~11105152       &  96  & 11.55  &  A* &$-2.2\pm1.4 ;           -0.1\pm1.4$&  48 &    -                              &  -          &$-23\pm4\ / -15\pm1$&70/79/93    \\
\hline
\end{tabular}							
}%}								
%\end{minipage}
\end{center}
\end{table*}
%Rv (HD123224=-13.89+/-2.10; HD123202=-15.02; HD123096=-8.0+/-1.4)

%% file: Table_Parameters.tex
\begin{table}
\caption[ ]{Atmospheric parameters for the programme stars. The last six entries 
refer to SB2 stars for which only very few parameters could be estimated, and
no abundance analysis was performed. The \logg\ values given for the two components of 
HD~123225 are assumed and not measured, therefore it is not possible to give a
realistic uncertainty to \Teff\ as well. Due to the very few measurable spectral 
lines, \vmic\ for HD~123202 is assumed.}
\label{Tab_Parameters}
\centering                      
\begin{tabular}{lr@{\,$\pm$\,}lr@{\,$\pm$\,}lr@{\,$\pm$\,}lr@{\,$\pm$\,}l}
\hline
\hline
Star &\multicolumn{2}{c}{\Teff}&\multicolumn{2}{c}{\logg} & \multicolumn{2}{c}{\vmic}  &\multicolumn{2}{c}{\vsini} \\ 
name &\multicolumn{2}{c}{[K]}  &\multicolumn{2}{c}{[cgs]} & \multicolumn{2}{c}{[\kms]} &\multicolumn{2}{c}{[\kms]} \\
\hline											
HD~122983            & 10700 & 200 & 4.00 & 0.15 & 1.0 & 0.2 &  35 &  3  \\
HD~123097            & 12000 & 300 & 3.96 & 0.10 & 1.5 & 0.6 & 133 &  5  \\   
HD~123182            & 10800 & 200 & 4.25 & 0.15 & 0.0 & 0.4 &  81 &  6  \\
HD~123183            & 10900 & 250 & 4.05 & 0.15 & 0.6 & 0.8 & 275 & 14  \\    
HD~123184            & 10900 & 250 & 4.25 & 0.15 & 0.0 & 0.4 &  60 &  6  \\     
HD~123201B           & 11700 & 250 & 3.95 & 0.15 & 0.0 & 0.8 & 202 & 11  \\    
HD~123202            & 11400 & 250 & 3.70 & 0.15 & 0.0 & 1.0 & 301 & 17  \\    
HD~123226            & 12400 & 300 & 4.04 & 0.10 & 0.0 & 0.2 &  17 &  1  \\  
HD~123269            & 11600 & 250 & 3.94 & 0.15 & 0.0 & 0.2 &  25 &  2  \\   
CD-47~8868           & 10400 & 200 & 4.20 & 0.15 & 0.9 & 0.7 & 250 & 10  \\
CD-47~8869	     &  8400 & 200 & 3.60 & 0.20 & 1.2 & 0.7 & 230 & 10  \\    
CD-47~8889           & 10300 & 200 & 4.15 & 0.15 & 0.5 & 0.6 & 113 &  3  \\
CD-47~8905	     &  9700 & 200 & 4.10 & 0.20 & 1.6 & 0.4 & 160 &  6  \\
CPD-47~6379          & 11000 & 250 & 4.30 & 0.15 & 1.7 & 0.6 & 136 &  7  \\
CPD-47~6385          &  9350 & 200 & 4.12 & 0.20 & 1.6 & 0.3 &  55 &  5  \\
UCAC~11104969        &  7050 & 150 & 4.00 & 0.20 & 1.4 & 0.9 & 260 & 15  \\  
UCAC~11105038        &  7650 & 200 & 4.00 & 0.20 & 3.4 & 0.4 &  85 &  5  \\  
UCAC~11105106        &  7500 & 200 & 4.00 & 0.20 & 2.5 & 0.6 & 195 & 10  \\  
UCAC~11105176        &  8100 & 200 & 4.00 & 0.20 & 1.8 & 0.4 & 125 &  7  \\   
UCAC~11105213        &  7600 & 200 & 4.00 & 0.20 & 1.9 & 0.3 &  57 &  3  \\    
UCAC~11105379        &  7700 & 200 & 4.00 & 0.20 & 2.1 & 0.2 &  44 &  2  \\  [2mm]
\hline
HD~123225A            & 13100 & -   & 3.8 &  -    &  &   & 12 &  1   \\
HD~123225B            & 8000  & -   & 4.0 &  -    &  &   & 20 &  2   \\
NGC~5460~BB\,338A & \multicolumn{6}{c}{-}            &140 & 7   \\
NGC~5460~BB\,338B & \multicolumn{6}{c}{-}            &$<$11& 1   \\
UCAC~11105152A       & \multicolumn{6}{c}{-}            & 190 & 10   \\
UCAC~11105152B       & \multicolumn{6}{c}{-}            &$<11$& 1   \\
\hline								 
\end{tabular}							 
\end{table}

%% file: Table_Abundances.tex
\begin{table*}
\caption[ ]{Element abundances, in $\log(N_{X}/N_{\rm tot})$, of the analysed cluster member
stars. The estimated internal errors in units of 0.01\,dex and the number of lines selected 
for each element are given in parenthesis. For comparison, the last row shows the solar 
abundances by \citet{met05}. Abundances obtained from just one line have no error (-;1).
The column labelled as `Uncertainty' shows the abundance uncertainty for each star, calculated following 
\citet{fossati2008}, and which is taken into account in the discussion of the results.}
\label{ngc5460 abn memb}
%\rotatebox{90}{
\scriptsize{
\begin{tabular}{lcccccccc}
\hline
\hline
Star & He & C & O & Na & Mg & Al & Si & S \\
\hline
HD~122983     & -2.00(-;1)    & 	      & -3.41(-;1)    & 	      & -5.22(-;1)    & -6.19(-;1)    & -4.41(-;1)    & 	      \\
HD~123097     & -0.90(-;1)    & 	      & 	      & 	      & -4.55(36;3)   & -5.85(-;1)    & -4.54(01;2)   & 	      \\
HD~123182     & -1.48(-;1)    & 	      & -3.41(-;1)    & 	      & -4.56(01;2)   & -5.43(-;1)    & -4.58(09;2)   & 	      \\
HD~123183     & 	      &  	      & -3.40(11;4)   & 	      & -4.30(16;3)   & 	      & -4.37(19;2)   & 	      \\
HD~123184     & -0.90(05;3)   & -3.47(-;1)    & -3.28(01;3)   & 	      & -4.35(16;5)   & -5.56(-;1)    & -4.39(10;4)   & 	      \\
HD~123201B    & -1.00(03;6)   & -3.60(-;1)    & -3.28(-;1)    & 	      & -4.29(10;5)   & -5.70(-;1)    & -4.52(01;2)   & 	      \\
HD~123202     &	              & 	      & 	      & 	      & 	      & 	      & -4.62(-;1)    & 	      \\
HD~123226     & -1.02(05;2)   & 	      & -3.41(01;2)   & 	      & -4.64(08;3)   & -5.77(17;2)   & -4.63(-;1)    & -4.77(17;6)   \\
HD~123269     & -0.98(-;1)    & 	      & 	      & 	      & -4.58(16;3)   & -5.76(-;1)    & -4.42(-;1)    & -4.45(-;1)    \\
CD-47~8868    & 	      & 	      & 	      & 	      & -3.80(-;1)    & 	      & -4.52(-;1)    & 	      \\
CD-47~8869    & 	      & -3.50(-;1)    & 	      & 	      & -5.27(16;2)   & 	      & -4.34(-;1)    & 	      \\
CD-47~8889    & 	      & 	      & 	      & 	      & -4.20(06;3)   & -5.52(-;1)    & -4.52(17;2)   & 	      \\
CD-47~8905    & 	      & 	      & 	      & 	      & -4.03(11;3)   & 	      & -4.26(09;2)   & 	      \\
CPD-47~6379   & 	      & 	      & 	      & 	      & -4.10(28;3)   & 	      & -4.77(45;2)   & 	      \\
CPD-47~6385   & 	      & -3.27(13;5)   & 	      & 	      & -4.28(02;3)   & -5.35(-;1)    & -4.71(24;2)   & 	      \\
UCAC~11104969 & 	      & -3.47(-;1)    & 	      & 	      & -5.11(32;2)   & 	      & -3.58(-;1)    & 	      \\
UCAC~11105038 & 	      & -3.38(14;3)   & 	      & -5.78(03;2)   & -4.86(07;3)   & 	      & -4.37(15;3)   & 	      \\
UCAC~11105106 & 	      & 	      & 	      & -6.13(06;2)   & -4.62(21;4)   & 	      & -4.09(21;2)   & 	      \\
UCAC~11105176 & 	      & -3.24(04;2)   & 	      & 	      & -4.45(03;3)   & 	      & -4.21(-;1)    & 	      \\
UCAC~11105213 & 	      & -3.49(12;3)   & 	      & -5.69(04;2)   & -4.87(06;3)   & 	      & -4.27(13;3)   & -4.37(-;1)    \\
UCAC~11105379 & 	      & -3.42(05;3)   & 	      & -5.84(05;2)   & -4.69(03;3)   & 	      & -4.37(09;3)   & -4.58(-;1)    \\
sun           & -1.11         & -3.65         & -3.38         & -5.87         & -4.51         & -5.67         & -4.53         & -4.90         \\
\hline
Star & Ca & Sc & Ti & V & Cr & Mn & Fe & Ni \\
\hline
HD~122983     & 	      & -7.48(20;4)   & -7.37(13;3)   & 	      & -5.27(14;19)  & 	      & -3.81(12;100) & 	      \\
HD~123097     & 	      & 	      & -7.31(10;2)   & 	      & -6.78(12;2)   & 	      & -4.60(05;12)  & 	      \\
HD~123182     & -5.29(-;1)    & 	      & -6.87(08;5)   & 	      & -5.90(23;9)   & -5.19(23;9)   & -4.20(14;60)  & 	      \\
HD~123183     & 	      & 	      & -6.51(-;1)    & 	      & -5.81(01;2)   & 	      & -4.37(22;11)  & 	      \\
HD~123184     & -5.77(-;1)    & -9.21(-;1)    & -6.88(14;8)   & 	      & -6.15(13;10)  & -6.10(-;1)    & -4.29(14;57)  & 	      \\
HD~123201B    & 	      & 	      & -6.54(18;5)   & 	      & -5.99(14;4)   & 	      & -4.47(14;20)  & 	      \\
HD~123202     &		      & 	      & 	      & 	      & -5.94(10;2)   & 	      & -4.50(07;5)   & 	      \\
HD~123226     & 	      & 	      & -7.27(-;1)    & 	      & -6.63(12;6)   & 	      & -4.75(15;54)  & -5.86(-;1)    \\
HD~123269     & 	      & 	      & -7.26(25;2)   & 	      & -6.48(13;9)   & 	      & -4.57(16;60)  & -5.76(-;1)    \\
CD-47~8868    & 	      & 	      & -7.04(15;2)   & 	      & -6.40(25;4)   & 	      & -4.25(21;13)  & 	      \\
CD-47~8869    & -5.35(-;1)    & -9.68(-;1)    & -6.67(08;2)   & 	      & -6.58(11;4)   & -6.82(-;1)    & -4.50(10;15)  & 	      \\
CD-47~8889    & 	      & 	      & -7.07(13;3)   & 	      & -6.16(09;5)   & 	      & -4.27(17;22)  & 	      \\
CD-47~8905    & 	      & 	      & -7.00(28;4)   & 	      & -6.39(10;4)   & 	      & -4.43(15;21)  & 	      \\
CPD-47~6379   & 	      & 	      & -7.12(22;4)   & 	      & -5.15(15;4)   & 	      & -4.20(23;27)  & 	      \\
CPD-47~6385   & -5.30(09;4)   & -8.81(04;3)   & -6.85(13;12)  & 	      & -6.04(09;17)  & -5.99(03;2)   & -4.38(09;59)  & -5.55(30;2)   \\
UCAC~11104969 & -5.69(17;4)   & -9.43(-;1)    & -7.48(32;5)   & 	      & -6.76(15;6)   & -6.34(08;3)   & -4.62(05;12)  & -5.65(65;5)   \\
UCAC~11105038 & -6.23(09;8)   & -9.17(19;3)   & -7.64(17;19)  & 	      & -6.50(15;16)  & -6.83(14;3)   & -5.08(13;91)  & -6.09(18;9)   \\
UCAC~11105106 & -5.67(07;8)   & -8.56(22;2)   & -7.44(14;10)  & 	      & -6.81(09;10)  & -6.83(08;3)   & -5.03(15;54)  & -6.04(04;3)   \\
UCAC~11105176 & -5.86(06;4)   & -8.82(11;2)   & -6.97(19;7)   & 	      & -6.33(08;10)  & -6.35(17;3)   & -4.80(12;61)  & -5.74(16;5)   \\
UCAC~11105213 & -6.03(25;19)  & -9.23(09;4)   & -7.16(04;13)  & 	      & -6.27(15;24)  & -6.75(22;5)   & -4.61(16;133) & -5.67(16;15)  \\
UCAC~11105379 & -5.89(14;15)  & -9.11(11;4)   & -7.22(05;15)  & -7.76(-;1)    & -6.39(12;22)  & -6.78(12;4)   & -4.74(16;129) & -6.11(18;14)  \\
sun           & -5.73	      & -8.99	      & -7.14	      & -8.04	      & -6.40	      & -6.65	      & -4.59	      & -5.81	      \\
\hline
Star & Cu & Zn & Sr & Y & Ba & Pr & Nd & Uncertainty \\
\hline
HD~122983     & 	      & 	      & 	      & 	      & -8.64(-;1)    & -7.57(20;3)   & -7.43(10;8)   & 0.20	      \\
HD~123097     & 	      & 	      & 	      & 	      & 	      & 	      & 	      & 0.26	      \\
HD~123182     & 	      & 	      & 	      & -9.81(-;1)    & -9.09(-;1)    & 	      & 	      & 0.23	      \\
HD~123183     & 	      & 	      & 	      & 	      & 	      & 	      & 	      & 0.34	      \\
HD~123184     & 	      & 	      & -9.34(-;1)    & 	      & 	      & 	      & 	      & 0.22	      \\
HD~123201B    & 	      & 	      & 	      & 	      & 	      & 	      & 	      & 0.30	      \\
HD~123202     &		      & 	      & 	      & 	      & 	      & 	      & 	      & 0.36	      \\
HD~123226     & 	      & 	      & 	      & 	      & 	      & 	      & 	      & 0.20	      \\
HD~123269     & 	      & 	      & 	      & 	      & 	      & 	      & 	      & 0.20	      \\
CD-47~8868    & 	      & 	      & 	      & 	      & 	      & 	      & 	      & 0.33	      \\
CD-47~8869    & 	      & 	      & 	      & 	      & -9.70(-;1)    & 	      & 	      & 0.32	      \\
CD-47~8889    & 	      & 	      & 	      & 	      & 	      & 	      & 	      & 0.25	      \\
CD-47~8905    & 	      & 	      & 	      & 	      & 	      & 	      & 	      & 0.28	      \\
CPD-47~6379   & 	      & 	      & 	      & 	      & 	      & 	      & 	      & 0.26	      \\
CPD-47~6385   & 	      & 	      & 	      & 	      & -8.17(11;2)   & 	      & 	      & 0.21	      \\
UCAC~11104969 & 	      & 	      & 	      & -8.64(03;2)   & -8.55(-;1)    & 	      & 	      & 0.34	      \\
UCAC~11105038 & 	      & 	      & 	      & 	      & -9.18(16;2)   & 	      & 	      & 0.23	      \\
UCAC~11105106 & 	      & -6.79(-;1)    & 	      & 	      & -9.63(-;1)    & 	      & 	      & 0.30	      \\
UCAC~11105176 & 	      & 	      & 	      & 	      & -10.10(-;1)   & 	      & 	      & 0.26	      \\
UCAC~11105213 & 	      & -7.15(-;1)    & 	      & -9.64(16;3)   & -8.51(16;2)   & 	      & 	      & 0.22	      \\
UCAC~11105379 & 	      & -7.75(-;1)    & 	      & -9.87(13;2)   & -9.35(16;2)   & 	      & 	      & 0.21	      \\
sun           & -7.83         & -7.44         & -9.12         & -9.83         & -9.87         & -11.33        & -10.59        &               \\ 
\hline
\end{tabular}
}
%}
\end{table*}

%% file: Table_Errors.tex
\begin{table*}
\caption[ ]{Error sources for the abundances of the chemical elements of CPD-47\,6385 and CD-47\,8905. 
Both stars have very similar atmospheric parameters (\Teff$\sim$9500\,K, \logg$\sim$4.1 and \vmic$\sim$1.6\,\kms), 
but rather different \vsini\ values, as shown in the second table line. Columns 2$-$4 and 6$-$8 give the variation 
in abundance estimated by increasing \Teff\ by 250\,K, \logg\ by 0.1\,dex, and \vmic\ by 0.7\,\kms, respectively.
Columns 5 and 9 give the mean error calculated applying the standard error propagation theory on the 
systematic uncertainties given in Col.~2$-$4 and 6$-$8, i.e., 
$\sigma_{\rm abn}^2$\,(syst.)    = 
$\sigma_{\rm abn}^2$\,(\Teff)  + 
$\sigma_{\rm abn}^2$\,(\logg)  + 
$\sigma_{\rm abn}^2$\,(\vmic). }
\label{Tab_Errors}
\centering                      
\begin{small}
\begin{tabular}{l|cccc|cccc}
\hline
\hline
 & \multicolumn{4}{c}{CPD-47\,6385}    & \multicolumn{4}{c}{CD-47\,8905}      \\
 & \multicolumn{4}{c}{\vsini=55\,\kms} & \multicolumn{4}{c}{\vsini=160\,\kms} \\
\hline											
El. & $\sigma_{\rm abn}$(\Teff) & $\sigma_{\rm abn}$(\logg) & $\sigma_{\rm abn}$(\vmic) & $\sigma_{\rm abn}$(syst.) & $\sigma_{\rm abn}$(\Teff) & $\sigma_{\rm abn}$(\logg) & $\sigma_{\rm abn}$(\vmic) & $\sigma_{\rm abn}$(syst.) \\
    & (dex) & (dex) & (dex) & (dex) & (dex) & (dex) & (dex) & (dex) \\
\hline											
C  & 0.15 & $-$0.04 & $-$0.01 & 0.15 & & & & \\
Mg & 0.21 & $-$0.07 & $-$0.15 & 0.27 & 0.31 & $-$0.07 & $-$0.28 & 0.42 \\
Al & 0.00 &    0.06 & $-$0.03 & 0.07 & & & & \\ 
Si & 0.09 & $-$0.03 & $-$0.04 & 0.10 & 0.05 & $-$0.01 & $-$0.06 & 0.08 \\
Ca & 0.25 & $-$0.04 & $-$0.09 & 0.27 & & & & \\
Sc & 0.14 &    0.04 & $-$0.03 & 0.15 & & & & \\
Ti & 0.08 &    0.01 & $-$0.17 & 0.19 & 0.17 &    0.05 & $-$0.31 & 0.36 \\
Cr & 0.13 &    0.03 & $-$0.08 & 0.16 & 0.17 &    0.06 & $-$0.19 & 0.26 \\
Mn & 0.23 & $-$0.01 & $-$0.03 & 0.23 & & & & \\
Fe & 0.13 &    0.00 & $-$0.08 & 0.15 & 0.09 &    0.04 & $-$0.16 & 0.19 \\
Ni & 0.10 & $-$0.03 &    0.00 & 0.10 & & & & \\
Ba & 0.26 & $-$0.07 & $-$0.48 & 0.55 & & & & \\
\hline								 
\end{tabular}							 
\end{small}
\end{table*}

%% file: Table_Second_ABN.tex
\begin{table*}
\caption[ ]{Best fit parameters from the comparison analysis with {\sc Zeeman}.  
Chemical abundances are in units of $\log(N_{X}/N_{tot})$. 
Elements marked with an asterisk are based on less than $\sim3$ useful lines 
and have uncertainties estimated by eye.  
}
\label{comparison-abn-tab}
%\scriptsize{
\begin{tabular}{lccccccc}
\hline\hline
        & HD 123182        & HD 123183         & HD 123184         & HD 123201B        &	HD 123226         & CPD-476379        & CPD-476385        \\	
\hline
\Teff\ $[$K$]$	
        & $10800 \pm 300$  & $10500 \pm 700 $  & $10900 \pm 120 $  & $12000 \pm 450 $  &	$12140 \pm 160 $  & $10100 \pm 300 $  & $9360  \pm 150 $  \\                               
\logg\ $[$cgs$]$  
        & $4.25	 \pm 0.2$  & $4.3   \pm 0.3 $  & $4.35  \pm 0.2 $  & $4.2   \pm 0.3 $  &	$4.27  \pm 0.1 $  & $4.15  \pm 0.2 $  & $4.12  \pm 0.2 $  \\                      
\vsini\ $[$\kms$]$
        & $81	 \pm 4	$  & $270   \pm 15  $  & $63.1  \pm 1.5 $  & $203   \pm 6   $  &	$18.1  \pm 1.0 $  & $140   \pm 8   $  & $56.1  \pm 2.5 $  \\                            
\vmic\ $[$\kms$]$
        & $\sim0 	$  & $\sim0         $  & $0.4   \pm 0.22$  & $\sim0         $  &	$0.5   \pm 0.5 $  & $1.02  \pm 0.45$  & $1.8   \pm 0.2 $  \\                                
Vr\ $[$\kms$]$ 
        & $-1	 \pm 6	$  & $-18	  \pm 20    $ & $-14   \pm 1 $  & $-25  \pm 13     $   &	$-12   \pm 1 $  & $-30 \pm 3 $  & $-21 \pm 1      $  \\[3mm]
He      & $-1.48 \pm 0.4$ *&                   & $-0.9  \pm 0.2 $ *& $-1.05 \pm 0.2 $ *&	$-0.9  \pm 0.2 $ *& $	           $  & $	       $  \\
C       &                  &  	               & $-3.4  \pm 0.2 $ *& $	           $   &	$              $  & $	           $  & $	       $  \\
O       & $-3.54 \pm 0.1$  & $-3.5  \pm 0.2 $ *& $-3.3  \pm 0.12$  & $-3.45 \pm 0.2 $ *&	$-3.57 \pm 0.2 $ *& $-3.4  \pm 0.21$  & $-3.3  \pm 0.15$ *\\
Na      &                  &  	               &                   & $	           $   &	$              $  & $	           $  & $-5.48 \pm 0.2 $ *\\
Mg      & $-4.5	 \pm 0.2$  & $-4.9  \pm 0.4 $ *& $-4.56 \pm 0.14$  & $-4.49 \pm 0.24$  &	$-4.87 \pm 0.06$  & $-4.47 \pm 0.17$  & $-4.40 \pm 0.05$  \\
Al      & $-6.1	 \pm 0.5$ *&  	               & $-5.6  \pm 0.2 $ *& $-5.85 \pm 0.4 $ *&	$-5.65 \pm 0.15$  & $-5.68 \pm 0.3 $ *& $	       $  \\
Si      & $-4.7	 \pm 0.2$  & $-4.5  \pm 0.2 $ *& $-4.4  \pm 0.11$  & $-4.71 \pm 0.27$  &	$-4.57 \pm 0.08$  & $-4.85 \pm 0.15$ *& $-4.59 \pm 0.1 $ *\\
S       &                  &  	               & $-4.4  \pm 0.3 $ *& $	           $   &	$-4.92 \pm 0.3 $ *& $	           $  & $	       $  \\
Ca      & $-5.68 \pm 0.4$ *&  	               & $-5.52 \pm 0.27$  & $	           $   &	$              $  & $-5.67 \pm 0.3 $ *& $-5.45 \pm 0.15$  \\
Sc      &                  &  	               & $-8.8  \pm 0.1 $  & $	           $   &	$              $  & $-8.48 \pm 0.2 $ *& $-8.81 \pm 0.09$  \\
Ti      & $-6.88 \pm 0.1$  & $-7.3  \pm 0.4 $ *& $-7.16 \pm 0.12$  & $-7.26 \pm 0.15$  &	$-7.4  \pm 0.1 $  & $-6.87 \pm 0.08$  & $-6.88 \pm 0.06$  \\
V       &                  &  	               & $-8.4  \pm 0.4 $ *& $	           $   &	$              $  & $	           $  & $	       $  \\
Cr      & $-6.24 \pm 0.1$  & $-6.9  \pm 0.4 $ *& $-5.99 \pm 0.11$  & $-6.67 \pm 0.2 $  &	$-6.53 \pm 0.05$  & $-5.53 \pm 0.11$  & $-6.21 \pm 0.14$  \\
Mn      & $-5.52 \pm 0.1$  & 	               & $-6.4  \pm 0.3 $ *& $		   $   &	$-6.72 \pm 0.3 $ *& $	           $  & $	       $  \\
Fe      & $-4.31 \pm 0.1$  & $-4.4  \pm 0.2 $  & $-4.41 \pm 0.14$  & $-4.66 \pm 0.19$  &	$-4.73 \pm 0.05$  & $-4.16 \pm 0.05$  & $-4.35 \pm 0.15$  \\
Ni      & 		   &		       &                   &		       &			  &		      & $	       $  \\
Cu      & 		   &		       &                   &		       &			  &		      & $	       $  \\
Zn      & 		   &		       &                   &		       &			  &		      & $	       $  \\
Y       & 		   &		       & $-9.6  \pm 0.4 $ *&		       &			  &		      & $-9.19 \pm 0.4 $ *\\
Ba      & 		   &		       & $-9.6  \pm 0.4 $ *&		       &			  &		      & $-8.85 \pm 0.4 $ *\\
\hline
\end{tabular}
\end{table*}

%% file: Table_Luminosities.tex
\begin{table}
\caption[]{\logl, $\log$\Teff, M/M$_{\odot}$ and fractional age ($\tau$) with associated error bars 
for the stars of the NGC~5460 open cluster.}
\label{table luminosity-ngc5460}
\centering                      
\begin{footnotesize}
\begin{tabular}{lccr@{\,$\pm$\,}lr@{\,$\pm$\,}c}
\hline
\hline
Star & \logl & $\log$\Teff & \multicolumn{2}{c}{\M }& \multicolumn{2}{c}{$\tau$}   \\
\hline
HD~122983	      &  1.88  &    4.029 &    2.76 &	  0.14  &    0.37  &	0.15  \\
HD~123097	      &  2.25  &    4.079 &    3.40 &	  0.17  &    0.67  &	0.27  \\
HD~123182	      &  1.89  &    4.033 &    2.80 &	  0.14  &    0.38  &	0.16  \\
HD~123183	      &  1.80  &    4.037 &    2.73 &	  0.14  &    0.36  &	0.15  \\
HD~123184	      &  2.06  &    4.037 &    3.02 &	  0.21  &    0.48  &	0.19  \\
HD~123201B	      &  2.10  &    4.068 &    3.17 &	  0.16  &    0.55  &	0.22  \\
HD~123202	      &  2.33  &    4.057 &    3.45 &	  0.24  &    0.70  &	0.28  \\
%HD~123225A\#	      &  2.44  &    4.107 &    3.80 &	  0.27  &    0.92  &	0.37  \\
%HD~123225B\#	      &  2.11  &    3.903 &    2.80 &	  0.28  &    0.38  &	0.15  \\
HD~123226	      &  2.35  &    4.093 &    3.60 &	  0.18  &    0.79  &	0.32  \\
HD~123269	      &  2.11  &    4.064 &    3.15 &	  0.16  &    0.54  &	0.22  \\
CD-47~8868	      &  1.50  &    4.017 &    2.30 &	  0.16  &    0.22  &	0.09  \\
CD-47~8869	      &  1.25  &    3.924 &    1.94 &	  0.14  &    0.14  &	0.05  \\
%CD-47~8879A\#	      &  1.63  &    4.041 &    2.55 &	  0.18  &    0.29  &	0.12  \\
%CD-47~8879B\#	      &  1.46  &    3.929 &    2.15 &	  0.15  &    0.18  &	0.70  \\
CD-47~8889	      &  1.47  &    4.013 &    2.35 &	  0.16  &    0.23  &	0.09  \\
CD-47~8905	      &  1.41  &    3.987 &    2.20 &	  0.11  &    0.19  &	0.08  \\
CPD-47~6379	      &  1.76  &    4.041 &    2.71 &	  0.14  &    0.35  &	0.14  \\
CPD-47~6385	      &  1.52  &    3.971 &    2.27 &	  0.16  &    0.21  &	0.09  \\
%Cl NGC~5460~BB\,338A\#&  0.83  &    3.845 &    1.52 &	  0.11  &    0.07  &	0.03  \\
%Cl NGC~5460~BB\,338B\#&  0.83  &    3.845 &    1.52 &	  0.11  &    0.07  &	0.03  \\
UCAC~11104969	      &  0.72  &    3.848 &    1.48 &	  0.07  &    0.06  &	0.03  \\
UCAC~11105038	      &  0.92  &    3.884 &    1.66 &	  0.08  &    0.09  &	0.04  \\
UCAC~11105106	      &  0.88  &    3.875 &    1.60 &	  0.08  &    0.08  &	0.03  \\
%UCAC~11105152A\#      &  0.95  &    3.875 &    1.65 &	  0.12  &    0.09  &	0.03  \\
%UCAC~11105152B\#      &  0.95  &    3.875 &    1.65 &	  0.12  &    0.09  &	0.03  \\
UCAC~11105176	      &  1.06  &    3.908 &    1.78 &	  0.09  &    0.11  &	0.04  \\
UCAC~11105213	      &  0.80  &    3.881 &    1.59 &	  0.08  &    0.08  &	0.03  \\
UCAC~11105379	      &  0.88  &    3.886 &    1.64 &	  0.08  &    0.08  &	0.03  \\
\hline		      									
\end{tabular}	      									
\end{footnotesize}						 
\end{table}

%% file: Table_Correlations.tex
\begin{table}
\caption[ ]{Correlations and relative uncertainties of the Mg and Fe 
element abundance as a function of \Teff, in two temperature ranges: 
\Teff$<$10500\,K and \Teff$>$10500\,K. The last two rows show the correlations
obtained for Mg and Fe from the \citet{erspamer} data.}
\label{correlations}
\centering                      
\begin{tabular}{llcc}
\hline
\hline
Element & \Teff & slope & $\sigma_{slope}$ \\
\hline									
Mg & $<$10500\,K &    3.04$\times$10$^{-4}$ & 0.42$\times$10$^{-4}$ \\
Mg & $>$10500\,K & $-$1.71$\times$10$^{-4}$ & 1.08$\times$10$^{-4}$ \\
Fe & $<$10500\,K &    1.88$\times$10$^{-4}$ & 0.47$\times$10$^{-4}$ \\
Fe & $>$10500\,K & $-$4.03$\times$10$^{-4}$ & 0.82$\times$10$^{-4}$ \\
\hline								 
Mg & $<$10500\,K &    7.23$\times$10$^{-5}$ & 1.80$\times$10$^{-5}$ \\
Fe & $<$10500\,K &    1.21$\times$10$^{-4}$ & 0.21$\times$10$^{-4}$ \\
\hline								 
\end{tabular}							 
\end{table}